%% file: main.tex
\documentclass{IEEEtran4PSCC}
\usepackage{cite}
\ifCLASSINFOpdf
   \usepackage[pdftex]{graphicx}
\else
   \usepackage[dvips]{graphicx}
\fi
\usepackage{svg}

\usepackage[cmex10]{amsmath}
\usepackage[caption=false,font=footnotesize]{subfig}
\usepackage{adjustbox}
\usepackage{gensymb}
\usepackage{tabularx}
\usepackage{booktabs}
\usepackage{tikz}
\usetikzlibrary{shapes.geometric,arrows,positioning}
\usepackage{pgfplots}\pgfplotsset{compat=newest, set layers}\usepgfplotslibrary{statistics}
\usetikzlibrary{spy}
\usepackage{circuitikz}
\usepackage[hyphens]{url}
\usepackage{textcomp}
\usepackage{}

\newcolumntype{Y}{>{\centering\arraybackslash}X}

\definecolor{crimson2143940}{HTML}{960767}
\definecolor{darkorange25512714}{HTML}{e94649}
\definecolor{forestgreen4416044}{HTML}{ffa600}
\definecolor{steelblue31119180}{HTML}{10155c}

\definecolor{pscc_blue}{HTML}{6691ce}
\definecolor{pscc_purple}{HTML}{9565c9}
\definecolor{pscc_green}{HTML}{5ba962}
\definecolor{pscc_pink}{HTML}{c85990}
\definecolor{pscc_orange}{HTML}{cb5f46}
\definecolor{pscc_gold}{HTML}{ad973e}

\makeatletter
\let\old@ps@headings\ps@headings
\let\old@ps@IEEEtitlepagestyle\ps@IEEEtitlepagestyle
\def\psccfooter#1{%
    \def\ps@headings{%
        \old@ps@headings%
        \def\@oddfoot{\strut\hfill#1\hfill\strut}%
        \def\@evenfoot{\strut\hfill#1\hfill\strut}%
    }%
    \def\ps@IEEEtitlepagestyle{%
        \old@ps@IEEEtitlepagestyle%
        \def\@oddfoot{\strut\hfill#1\hfill\strut}%
        \def\@evenfoot{\strut\hfill#1\hfill\strut}%
    }%
    \ps@headings%
}
\makeatother

\psccfooter{%
        \parbox{\textwidth}{\hrulefill \\ \small{24th Power Systems Computation Conference} \hfill \begin{minipage}{0.2\textwidth}\centering \vspace*{4pt} \includegraphics[scale=0.06]{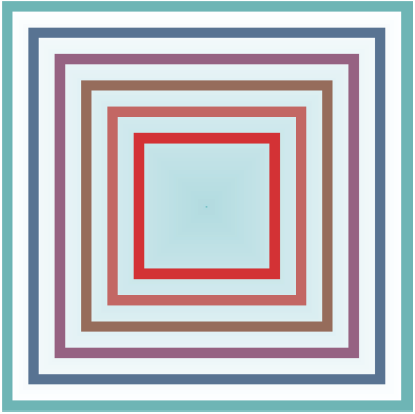}\\\small{PSCC 2026} \end{minipage} \hfill \small{Limassol, Cyprus --- June 8-12, 2026}}%
}

\begin{document}
\title{Risk-Based Dynamic Thermal Rating in Distribution Transformers via Probabilistic Forecasting}

%\title{Scalable Probabilistic Forecasting for Dynamic Thermal Ratings in Distribution Transformers}

%\title{Enabling Distribution Transformer\\Dynamic Thermal Ratings via\\Scalable Probabilistic Forecasting}

%\title{Risk-Based Operation of Distribution Transformers\\via Scalable Probabilistic Forecasting for enabling\\Automated Dynamic Thermal Rating}

% other possible key-words - risk-based, data-driven, machine learning

%% To specify the authors when (number of affiliations > 2)
\author{\IEEEauthorblockN{Scott Angus\IEEEauthorrefmark{1},
Jethro Browell\IEEEauthorrefmark{2},
David Greenwood\IEEEauthorrefmark{1} and
Matthew Deakin\IEEEauthorrefmark{1}}
\IEEEauthorblockA{\IEEEauthorrefmark{1}School of Engineering\\
Newcastle University,
Newcastle-upon-Tyne, UK\\ \{s.angus2, david.greenwood, matthew.deakin\}@newcastle.ac.uk}
\IEEEauthorblockA{\IEEEauthorrefmark{2}School of Mathematics and Statistics\\
University of Glasgow,
Glasgow, UK\\ jethro.browell@glasgow.ac.uk}
}

\maketitle

\begin{abstract}
\input{abstract}
\end{abstract}

\section{Introduction}
\input{introduction}
\section{Automated Dynamic Thermal Rating}\label{sec:automated_dtr}
\input{automated_dtr_management}
\section{Thermal Modelling}\label{sec:thermal_modelling}
\input{thermal_modelling}
\section{Protection Modelling}\label{sec:relay_optimisation}
\input{relay_optimisation}
\section{Probabilistic Forecasting Methodology}\label{sec:methodology}
\input{methodology}
\section{Results}\label{sec:results}
\input{results}
\section{Conclusions}\label{sec:conclusions}
\input{conclusions}

\bibliographystyle{IEEEtran}

\bibliography{PSCC_bib}{}

\end{document}

%% file: abstract.tex
Low voltage (LV) distribution transformers face accelerating demand growth while replacement lead times and costs continue to rise, making improved utilisation of existing assets essential. Static and conservative protection devices (PDs) in distribution transformers are inflexible and limit the available headroom of the transformer.
This paper presents a probabilistic framework for dynamically forecasting optimal thermal protection settings.
The proposed approach directly predicts the day-ahead scale factor which maximises the dynamic thermal rating of the transformer from historical load, temperature, and metadata using clustered quantile regression models trained on 644 UK LV transformers. Probabilistic forecasting quantifies overheating risk directly through the prediction percentile, enabling risk-informed operational decisions.
Results show a 10--12\% additional capacity gain compared to static settings, with hotspot temperature risk matching the selected percentile, including under realistic temperature forecast errors.
These results demonstrate a practical approach for distribution network operators to take advantage of PDs with adaptive settings to maximise capacity and manage risk on operational time scales.

%% file: introduction.tex
By 2050, low voltage (LV) demand is forecast to increase by 50--100\%~\cite{spenergynetworks-2024}. Transformer lead times~\cite{dempsey-2024} and costs are also increasing~\cite{skidmore-2024}, limiting upgrades as a timely option. Static LV protection can be excessively conservative, constraining capacity even during safe conditions. As a result, transformer capacity is increasingly limited by protection policy rather than thermal limits.
This motivates the need for more flexible use of existing transformers. Dynamic thermal rating (DTR) allows some relaxation of constraints by adjusting transformer loading limits based on real-time or forecasted conditions~\cite{gao-2016}. While DTR has been widely studied at higher voltage levels, LV transformers rarely benefit due to limited monitoring and simple protection devices (PDs). 

Numerical relays have adjustable tripping thresholds which could help facilitate DTR, however to be operationally viable, Distribution Network Operators (DNOs) need a controllable method to balance overheating risk and capacity gain. Increasing the relay’s scale factor increases its tripping threshold, enabling higher peak loads via {DTR}.

Many innovation projects have focused on improving the flexibility of {LV} networks, making this a timely area of research as many of these require adaptable {PDs}. Meshing networks~\cite{northernirelandelectricitynetworks-2022} and deploying smart transformers with advanced PDs~\cite{spenergynetworks-2018} have been investigated as well as load sharing and solid-state switching initiatives~\cite{ukpowernetworks-2016,ukpowernetworks-2014}. These projects all demonstrated the potential of advanced {PDs} to enhance {LV} network flexibility and the motivation of DNOs to explore protection upgrades, highlighting the likelihood of their increased deployment in future {LV} networks to help provide additional control over static protection such as fuses and miniature circuit breakers. Notably, numerical relays can often be retrofitted without major substation modifications~\cite{lloyd-2025}, making such upgrades practical.
Existing research on relay setting optimisation focuses on applications such as microgrid protection~\cite{laaksonen-2014}, different network states~\cite{samadi-2020}, and coordinating distributed energy resources~\cite{el-khattam-2009}. This typically involves defining multiple setting groups and selecting the appropriate group based on network conditions. While effective for fault protection and coordination, this strategy lacks specificity to transformer thermal conditions and does not address the challenge of maximising transformer utilisation via DTR.

In parallel, load forecasting has become a key tool in power systems, with methods studied for LV networks~\cite{mamun-2020}. These are generally deterministic and aimed at supply-demand management. Forecasting has also been applied to transformer condition monitoring~\cite{doolgindachbaporn-2022}, however, using either of these approaches to derive relay settings introduces error propagation.

Reference~\cite{summers-2022} used deterministic, single transformer models to coordinate relays directly. We instead propose a probabilistic approach using clustered models aiming to maximise {LV} transformer capacity under hotspot constraints.

Overall, existing methods either focus on fault protection or rely on deterministic forecasts, limiting their ability to robustly optimise relay settings for thermal protection. Using deterministic approaches for DTR propagates uncertainty. This motivates the need for a probabilistic, risk-based approach that can directly forecast optimal relay settings to maximise DTR while quantifying the risk of transformer overheating. 

This work fills a literature gap by proposing a probabilistic, scalable method for directly forecasting optimal relay scale factors for {LV} transformers. This scale factor controls the tripping threshold of modern numerical relays, allowing for greater loading in advantageous conditions.

The contribution of this work is a novel risk-based methodology for assigning day-ahead relay settings that maximise transformer capacity for DTR whilst controlling the level of risk of unacceptable hotspot development.
We show that the trade-off between capacity gain and overload risk can be accurately controlled by the DNO via the choice of prediction percentile. This gives DNOs a practical tool to set protection policies in line with their risk appetite.

The paper is organised as follows. Section~\ref{sec:automated_dtr} discusses the requirements for using adaptable PDs to apply DTR to transformers, before Sections~\ref{sec:thermal_modelling} and~\ref{sec:relay_optimisation} review the thermal and protection models to present the physical and engineering basis underpinning the proposed forecasting approach. Section~\ref{sec:methodology} details the methodology, including clustering, feature engineering, and model training, to show how fleet-wide information can be incorporated into a scalable and effective forecasting approach. Section~\ref{sec:results} presents the results and discussion to demonstrate the accuracy of the proposed models and compares clustering approaches. Section~\ref{sec:conclusions} concludes with implications for practice and directions for future research.

%% file: automated_dtr_management.tex
Current DTR approaches rely on SCADA systems and operator intervention when assets are excessively overloaded. It is not practical to assume operator intervention can be transferred to distribution transformer DTR due to the number of assets, with utilities often managing hundreds of thousands of transformers. To facilitate the widespread uptake of DTR to distribution transformers, a more automated approach is therefore required. 

We propose that a DNO will want to take action to mitigate overload when the winding temperature reaches some threshold (rather than, for example, tripping at an instantaneous power or current value). Most distribution transformers (as analysed in this work) lack direct hotspot or top-oil temperature measurements, therefore, a thermal model is required to estimate the impact of a transformer's load and thermal parameters, covered fully in Section~\ref{ss:thermal_model}.

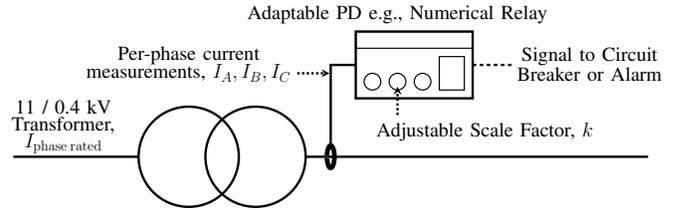
\begin{figure}
    \centering
    \resizebox{\columnwidth}{!}{%
        \centering
        \input{fig_Circuit_Diagram.tex}
    }
    \caption{The proposed adaptive transformer protection scheme considered to enable automated DTR. By choosing the PD's scale factor, \(k\), the current at which measured phase currents \(I_A,\,I_B,\,I_C\) trip can be adjusted, accounting for the impact of variable ambient temperature and off-peak load conditions on winding hotspot temperatures.}
    \label{fig:relay_diagram}
    \vspace{-0.5cm}
\end{figure}

Adaptable PDs, such as numerical relays, have the functionality to then automate the management of these transformers. An example implementation is shown in Fig.~\ref{fig:relay_diagram}, where the relay is located on the secondary side of the transformer as this is more suitable for {DTR} applications due to the preventative actions (such as feeder disconnects) being applied to the secondary side rather than the primary side as in fault protection approaches. This thermal protection (denoted as ANSI 49 in~\cite{ieee-2022}) can be provided via a dual time constant model (described in Section~\ref{ss:dual_tc_model}), which calculates the tripping current from the load and thermal parameters of the transformer, and is independent of the fault protection settings. The tripping current can be adjusted via the \emph{scale factor}, \(k\), to allow greater loading in advantageous conditions whilst controlling the risk of unacceptable hotspot temperatures. As the ANSI 49 function responds to current magnitudes rather than direction, reverse power flows from distributed energy resources do not affect the thermal protection settings.

A threshold temperature is required to determine the optimal scale factor which maximises the capacity gains from DTR whilst still providing sufficient protection from overheating (full details of this assignment is described in Section~\ref{ss:optimal_sf}).
We choose a temperature threshold of 140\(\degree\)C, as above this, there is increased likelihood of bubbles forming in the insulation, severely weakening the winding's dielectric strength.
It is used as the hotspot temperature limit for distribution transformers under normal cyclic loading in the IEC loading guide~\cite{iec-2018}, and is frequently used as the hotspot limit in both academic works~\cite{hajeforosh-2022,gao-2017} and industrial reports~\cite{gao-2016}.

The relay's true optimal scale factor can only be calculated retrospectively once the exact temperature and loads are known. Therefore, to enable the day-ahead transformer management, either implicit or explicit forecasting of an appropriate scale factor is required. To enable a DNO to have confidence in these forecasts at scale, we propose to use probabilistic forecasting approaches, which, if well-calibrated, enable effective and understandable risk-based asset management (selecting the \(p\)th percentile corresponds to a \(p\)\% probability of exceeding the hotspot limit).

In this work, we suggest to avoid the assignment of the scale factor indirectly from \emph{load} forecasts, which has two issues. Firstly, load forecasts are usually deterministic; probabilistic load forecasting remains an emerging topic~\cite{mamun-2020}. Deterministic methods require additional uncertainty quantification to enable risk-based decision-making. Secondly, calculating a scale factor from six forecasted quantities (three per-phase currents; peak and off peak conditions) compounds forecast errors, reducing accuracy (Section~\ref{sec:results}).

Instead, probabilistic forecasting is used to predicting the optimal relay scale factor directly (Section~\ref{sec:methodology}). By developing forecasting models trained using optimal scale factors calculated retrospectively, we can remove the reliance on intermediate load predictions, reducing the compounding errors, and provide interpretable risk metrics.

%% file: fig_Circuit_Diagram.tex
\begin{circuitikz}[american]
\usetikzlibrary{shapes.symbols}
\tikzstyle{every node}=[font=\footnotesize]
\draw [ line width=2pt](1,15.75) to[short] (4.75,15.75);
\draw [ line width=2pt ] (6.25,15.75) circle (1.5cm);
\draw [ line width=2pt ] (8.25,15.75) circle (1.5cm);
\draw [ line width=2pt](9.75,15.75) to[iloop] (11.25,15.75) -- (20,15.75);

% \draw [ line width=2pt ](0,0) -- ++ (1,0) to [iloop] ++(2,0);
\draw [ line width=2pt](10.5,18.5) to[short] (10.5,16);
% \draw[domain=10.47:12.19,samples=100,smooth, line width=2pt] plot (\x,{0.4*sin(7.375*\x r  + 1.25 r) +15.75});
\draw [ line width=2pt ] (11.25,19.5) rectangle (14.75,17.5);
\draw [ line width=1pt ] (11.25,19.5) rectangle (14.75,19);
\draw [ line width=1pt ] (12.5,18) circle (0.25cm);
\draw [ line width=1pt ] (11.75,18) circle (0.25cm);
\draw [ line width=1pt ] (13.25,18) circle (0.25cm);
\draw [ line width=1pt ] (13.75,18.75) rectangle (14.5,17.75);
\draw [ line width=2pt](10.466,18.5) to[short] (11.25,18.5);
% \draw [ line width=2pt](16.75,15.75) to[short] (18.25,15.75);
% \draw node[ line width=1.25pt, iecsocketL, rotate=-90, xscale=2.0, yscale=7] at (16, 16.12) {};
\draw [line width=2pt, dashed] (14.75,18.5) -- (16,18.5);
% \draw [line width=2pt, dashed] (16,18.5) -- (16,16.25);
\node [font=\LARGE] at (2.5,17.25) {11 / 0.4 kV};
\node [font=\LARGE] at (2.5,16.7) {Transformer,};
\node [font=\LARGE] at (2.5,16.15) {\(I_{\mathrm{phase\:rated}}\)};
\node [font=\LARGE, anchor = west] at (11.75,16.5) {Adjustable Scale Factor, \(k\)};
% \node [font=\LARGE] at (16,15) {Circuit Breaker};
\node [font=\LARGE] at (18.25,18.775) {Signal to Circuit};
\node [font=\LARGE] at (18.25,18.225) {Breaker or Alarm};
\node[font=\LARGE] at (12.5,20) {Adaptable PD e.g., Numerical Relay};
\draw [line width=2pt, dotted, ->, >=stealth] (12.5,17) -- (12.5,17.85);
\node[font=\LARGE] at (6.25,18.8) {Per-phase current};
\node[font=\LARGE] at (6.25,18.25) {measurements, \(I_A, I_B,I_C\)};
\draw [line width=2pt, dotted, ->, >=stealth] (9.5,18.25) -- (10.466,18.25);
% \node [font=\LARGE] at (6.75,19.25) {Adjustable};
% \node [font=\LARGE] at (6.75,18.5) {Scale Factor};

% \draw (0,0) -- ++(1,0) to[R] ++(3,0)
% to [iloop, mirror, name=I] ++(0,-2);
% \draw (1,0) to[oscope, v=$v$] ++(0,-2)
% node[ground]{};
% \draw (I.i) -- ++(-0.5,0) node[oscopeshape, anchor=right, name=O]{};
% \draw (O.south) -- (O.south |- GND) node[ground]{};
\end{circuitikz}

%% file: thermal_modelling.tex
\subsection{Basic Thermal Model}\label{ss:thermal_model}

The thermal model selected for our work is the IEEE thermal model~\cite{ieee-2012} which estimates the maximum hotspot temperature, \(\theta_H\), during the peak loading period as the sum of the ambient temperature, \(\theta_A\), top-oil rise above ambient, \(\Delta \theta_O\), and hotspot rise above top oil, \(\Delta \theta_H\),
\begin{equation}
  \label{eq:hs_sum}
  \theta_H = \theta_A + \Delta \theta_O+\Delta \theta_H\,.
\end{equation}

The full derivation of the model can be found in the IEEE standard~\cite{ieee-2012}, but the key parameter is the load factor (\(K\)), which is the ratio of the actual load to the rated load of the transformer, all other parameters being known or estimated from the transformer's nameplate data. 
This model uses a two-step equivalent load profile to estimate the top-oil and hotspot temperature rises. This is calculated using the RMS of the load during the 12 hours prior to the peak period, \(I_i\), and during the peak period itself, \(I_p\). The equivalent peak load is additionally constrained as a minimum of 90\% of the maximum half-hourly demand. The generic load factor, \(K_\mu\), is used to denote either \(K_i\) or \(K_p\) depending on the loading period.

\subsubsection{Accounting for Unbalanced Phase Loading}
The IEEE thermal model's assumption of a balanced load is not appropriate for distribution transformers with LV feeders connected at the secondary due to high levels of phase current unbalance. This is an important consideration as unbalanced phase loading with total equivalent power throughput can lead to significantly different hotspot temperatures~\cite{abdali-2024},~\cite{diao-2025}.

To account for this and to coordinate better with the protection device models, the load factor, \(K_\mu\), was split into a winding and top-oil component. The symbol \(\mu \) is used as a placeholder referring to either peak (\(p \)) or off-peak (\(i \)) quantities; e.g., \(K_{\mu} \) refers to either the off-peak load factor \(K_i \) or peak load factor \(K_p \). The winding load factor, \(K_{w,\:\mu}\), was calculated as the maximum per-phase load divided by the rated load,
\begin{equation}
    K_{w,\mu} = \frac{\max(I_{A,\mu}, I_{B,\mu}, I_{C,\mu})}{I_{\mathrm{phase\:rated}}}\,, \label{eq:k_winding}
\end{equation}
while the top-oil load factor (\(K_{o,\:\mu}\)) was calculated as the average per-phase load divided by the rated phase load,
\begin{equation}
    K_{o,\mu} = \frac{I_{\mathrm{mean}, \mu}}{I_{\mathrm{phase\:rated}}}\,. \label{eq:k_oil}
\end{equation}
This follows the approach used in~\cite{kong-2020} and accounts for the localized nature of hotspot formation.

The overall hotspot temperature, \(\theta_H\), can be described as a function of the peak and off-peak load factors as well as the ambient temperature, \(\theta_A\), and thermal parameters, \(\rho\), of the transformer. This can be expressed as:
\begin{equation}
    \theta_H = f_{\mathrm{IEEE}}(K_{o,i}, K_{o,p}, K_{w,i}, K_{w,p}, \theta_A, \rho)
\end{equation}
where Eq.~\eqref{eq:k_winding} and Eq.~\eqref{eq:k_oil} are used to calculate the winding and top-oil load factors, respectively.

Although we have used the IEEE model in this work for its optimisation simplicity and speed of implementation across a fleet of transformers, the proposed forecasting approach and scale factor optimisation method themselves are model agnostic and could be replaced by a black-box model or other models if desired. Lower temperature values or ageing models could also be used as alternative thresholds depending on the application and requirements of the DNO. The selection of thermal model is a trade-off between accuracy and complexity, and the best choice will depend on the specific application and available data. Comparison of these different models is beyond the scope of this work.

%% file: relay_optimisation.tex
\subsection{Dual Time Constant Model}\label{ss:dual_tc_model}

The scale factor, \(k\), scales the transformer rated phase current, \(I_{\mathrm{phase\:rated}}\), to give a scaled rated phase current, \(\hat{I}\),
\begin{equation}\label{eq:scaled_I_rated}
    \hat{I} = k \times I_{\mathrm{phase\:rated}}\,.
\end{equation}
The dual time constant model used in the relay uses this scaled phase rating, \(\hat{I}\), to calculate the tripping current, \(I_{\mathrm{trip}}\). This is based on the time constant of the winding, \(\tau_w\), and the insulation oil, \(\tau_o\), and is given by~\cite{gegridsolutions-2022}:
\begin{equation}
  \label{eq:relay_trip_current}
  I_{\mathrm{trip}}=\sqrt{\frac{0.4I_{i}^2e^{(-t/\tau_w)}+0.6I_{i}^2e^{(-t/\tau_o)}-\hat{I}^2}{0.4e^{(-t/\tau_w)}+0.6e^{(-t/\tau_o)}-1}}\,,
\end{equation}
where \(t\) is the time to trip, and the off-peak load is \(I_{i}\).

\subsection{Retrospective Scale Factor Calculation}\label{ss:optimal_sf}
The optimal scale factor allows the greatest tripping current, \(I_{\mathrm{trip}}\), such that the hotspot temperature, \(\theta_H\), does not exceed 140\(\degree\)C.
In order to find the maximum hotspot temperature for a given relay scale factor, the tripping current, \(I_{\mathrm{trip}}\), is used as the maximum per-phase load, with the other phase loads scaled proportionally to the peak phase increase. The hotspot temperature is then calculated using the thermal model described in Section~\ref{sec:thermal_modelling}. An iterative line-search algorithm (the bounded Brent method) was then used to find the value of \(k\) which gave the 140\(\degree\)C hotspot condition accurate to 0.01\(\degree\)C.

For each day and transformer, the ambient temperature was found using historical weather data and the optimal scale factor calculated retrospectively using the load monitoring data. This scale factor which maximises the loading without breaking the temperature threshold was used as the target function for the forecasting models described in Section~\ref{sec:methodology}. The scale factor is set once per day as sub-daily variation in loading is inherently captured through the dual-time constant model. By selecting a scale factor sufficiently conservative for the peak loading, we ensure that the transformer is also protected during lower off-peak loading periods.
Note that this optimal scale factor also provides an upper bound on the unlockable capacity for DTR without allowing any overheating events, as it assumes perfect knowledge of the load and temperature during the peak period.

%% file: methodology.tex
To realise well-calibrated probabilistic forecasts, we propose a three-step process: clustering, training, and forecasting. Clustering is beneficial as it overcomes the issue of data scarcity. In particular, long-term demand changes mean that monitoring data at a given transformer will become stale, so it is likely to be unrealistic that there would be more than a few hundred valid datapoints per transformer. A single forecasting model per-transformer would therefore lead to a high risk of overfitting and poor generalization; in contrast, a single global forecasting model for all transformers would not capture the variability between the transformers within the fleet. Clustering allows a DNO to aggregate the data of similar transformers based on their features, improving the generalization of the models while still capturing the specific characteristics of each group. 

\subsection{Clustering Approaches}

The raw features used for clustering are historic load and transformer metadata (see Section~\ref{sss:feature_engineering}). This raw data is pre-processed, then the K-means algorithm is used to determine the clusters from the pre-processed data. The optimal number of clusters is determined using combined silhouette and Bayes Information Criterion scores. 

To ensure good clustering performance, four pre-processing approaches were studied. One of two algorithms first processed the data (ridge regression or normalisation), then the resultant parameter weights or features (respectively), were either passed directly to the K-means function, or, via a principal component analysis (PCA) algorithm using the component weights. For the ridge regression approach, a separate model was trained for each transformer, whilst the normalisation approach scaled each feature to zero mean and unit variance across all transformers. For PCA, components which explained 90\% of the variance were used as features for clustering.

To determine the best performing clustering approach, models were trained on each cluster for each approach (Section~\ref{ss:training}) and evaluated on a holdout dataset. The clustering approach resulting in the best performance on the holdout set (Section~\ref{sss:quantifying_performance}) was selected for the final model.

\subsection{Training}\label{ss:training}
After assigning each transformer to a cluster, their aggregated data were used for feature engineering and model training. The dataset included load, calendar, and metadata features. Random noise was added initially, and features with lower importance were removed to reduce over-fitting.

\subsubsection{Feature engineering}\label{sss:feature_engineering}
The load data consisted of the mean per-phase loads during the peak and off-peak periods lagged by both a day and a week. An unbalance metric (the ratio of the range of phase currents to the mean) during the same periods was included as well as the daily and weekly lagged optimal scale factor.

The calendar data included the day of the week, weekend and holiday flags, and the  temperature during the peak period. An exponential weighted moving average of the temperature was also included to give some inertia to the change in demand due to temperature changes. A smoothing factor of 0.05 was used after testing values between 0.01 and 0.8 to find the best performing value when tested on the validation set.

The metadata used in the training dataset included the transformer rated power, the number of customers connected, and the transformer ID. This was included as a categorical variable to enable any unique characteristics of each transformer that may not be captured by other features to be learned.

\subsubsection{Single- and Multi-Temperature Training}
To investigate the impact of accurate temperature training data, two scenarios were considered. The first assumed that the exact temperature at the site of the transformer was known retrospectively, e.g. via a temperature sensor, and therefore a single temperature value was used for each day in the training period.

The second scenario used multiple temperature values within 2\(\degree\)C of the true value for each day in the training period. This was done in an attempt to reduce overfitting to the exact temperature values seen during training and make the model more resilient to weather forecast errors in the prediction stage. 

\subsection{Probabilistic Forecasting}
The forecasting algorithm selected was LightGBM~\cite{ke-2017} due to its support of quantile regression and efficiency. Quantile regression produces probabilistic outputs that map naturally to operational risk levels. The model parameters were optimised via Bayesian optimisation using Optuna~\cite{akiba-2019} to minimise the pinball loss on the validation set. A learning rate of 0.026 and 19 leaves were found to be optimal, with feature and bagging fractions of 0.93 and 0.87 respectively.
For each cluster, a separate LightGBM model was trained for each percentile investigated using the training dataset, and then incrementally retrained after each prediction. 

\subsubsection{Quantifying Probabilistic Forecast Performance}\label{sss:quantifying_performance}
To assess the reliability of the probabilistic forecast to support risk-based calculations, we assess forecast performance with the 5\% to 95\% \emph{coverage} rate, i.e., the model's calibration. The coverage returns the percentage of days when the true scale factor was within the 5th and 95th percentile predictions. A perfect model would have a coverage of 90\%, whereas an under-confident model would have greater coverage and an over-confident model would have less coverage.

\subsubsection{Clean and Noisy Temperature Prediction}
Two test datasets were used in the prediction stage, one assumed perfect weather forecasts and one used realistic weather forecasts. This was done to investigate the effect of forecast errors on the prediction accuracy. In the UK, 92.5\% of temperature forecasts are within 2\(\degree\)C~\cite{metoffice-2025} of the true temperature. To simulate this, normally distributed noise with a standard deviation of 1.12\(\degree\)C was added to the temperature values before making each prediction in a day-ahead fashion.

\subsubsection{Scale Factor Forecasting via Load Forecasting}\label{sss:multistage_method}

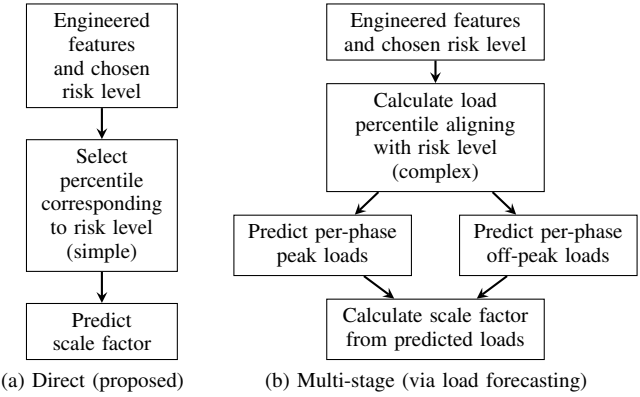
\begin{figure}
\centering
\subfloat[][Direct (proposed)\label{fig:direct_approach}]{%
    \centering
    \input{fig_flowchart_methodology_scale.tex}
}
\subfloat[][Multi-stage (via load forecasting)\label{fig:load_approach}]{%
    \centering
    \input{fig_flowchart_methodology_load.tex}
}
\caption{Flowcharts summarising the direct prediction of scale factors (a) and the calculation from load predictions (b).}
\label{fig:approach_comparison}
\vspace{-0.5cm}
\end{figure}

Fig.~\ref{fig:direct_approach} summarises the three steps for the proposed risk-based scale factor selection. As shown, this approach bypasses load forecasting, and directly forecasts the scale factor for the numerical relays using the retrospective optimal scale factors.

However, a DNO may already have some load forecasting capabilities and so may wish to adapt those load forecasts to achieve the same functionality as the proposed direct approach (as discussed in Section~\ref{sec:automated_dtr}). This would require additional steps to capture the impact of the load on the scale factor. An approach to achieve this is shown in Fig.~\ref{fig:load_approach}. To realise this multi-stage model, a collection of load prediction models were also trained using the same features and clustering as the direct approach. Each model was trained to predict the per-phase load during the peak or off-peak periods, resulting in six models per cluster. Using these predicted loads, the optimal scale factor was calculated using the method described in Section~\ref{sec:relay_optimisation}. Note that the increased number of models per cluster and calculations of correlations between phases increases the computational complexity.

This multi-stage approach becomes more complex for percentile predictions since per-phase loads are not independent. When calculating the scale factor from predicted loads, the 5th percentile factor was derived from the 95th percentile load predictions, and vice versa (as in Section~\ref{sec:relay_optimisation}). This yields a conservative estimate, as simultaneous 95th percentile loads across phases are unlikely. Although assuming perfect phase correlation is unrealistic, it offers a practical counterfactual and illustrates the method’s relative simplicity.

%% file: fig_flowchart_methodology_scale.tex
\tikzstyle{arrow} = [thick,->,>=stealth]
\tikzstyle{startstop} = [rectangle, rounded corners, text centered, draw=black, fill=none,]
\tikzstyle{process} = [rectangle, text centered, draw=black, fill=none,]
\newcommand{\blockwith}{0.2\columnwidth}
\begin{tikzpicture}[node distance=0.413cm]
    \node [font=\footnotesize, text width=\blockwith] (start) [process] {Engineered features and chosen risk level};
    \node [font=\footnotesize, text width=\blockwith, below=of start] (percentile) [process] {Select percentile corresponding to risk level (simple)};
    \node [font=\footnotesize, text width=\blockwith, below=of percentile] (predict) [process] {Predict scale factor};
    \draw [arrow] (start) -- (percentile);
    \draw [arrow] (percentile) -- (predict);
\end{tikzpicture}

%% file: fig_flowchart_methodology_load.tex
\tikzstyle{arrow} = [thick,->,>=stealth]
\tikzstyle{startstop} = [rectangle, rounded corners, text centered, draw=black, fill=none]
\tikzstyle{process} = [rectangle, text centered, draw=black, fill=none]
\newcommand{\blockwith}{0.3\columnwidth}
\begin{tikzpicture}[node distance=0.3cm]
    \node [process, font=\footnotesize, text width=\blockwith] (start) {Engineered features and chosen risk level};
    \node [process, font=\footnotesize, text width=\blockwith, below=of start] (percentile) {Calculate load percentile aligning with risk level (complex)};
    \node [process, font=\footnotesize, text width=0.24\columnwidth, below=of percentile, xshift=-1.5cm] (peak) {Predict per-phase peak loads};
    \node [process, font=\footnotesize, text width=0.24\columnwidth, below=of percentile, xshift=1.5cm] (off-peak) {Predict per-phase off-peak loads};
    \node [process, font=\footnotesize, text width=\blockwith, below=of off-peak, xshift=-1.5cm] (scale-factor) {Calculate scale factor from predicted loads};
    \draw [arrow] (start) -- (percentile);
    \draw [arrow] (percentile) -- (peak);
    \draw [arrow] (percentile) -- (off-peak);
    \draw [arrow] (peak) -- (scale-factor);
    \draw [arrow] (off-peak) -- (scale-factor);
\end{tikzpicture}

%% file: results.tex
The following section presents validation of the proposed method using per-phase current magnitude data from a fleet of 644 utility owned distribution transformers in the UK with capacities ranging from 25--1000 kVA, for 183 days through the winter of 2024/25~\cite{nationalgridLVLoadMonitor2024}. The data was then split into training and validation (September to January, 152 days) and holdout sets (February to early March, 31 days), capturing the winter peak period when capacity gains from DTR are greatest (due to correlated low ambient temperatures and peak demand). Weather data at the transformer sites was obtained from~\cite{open-meteo-2025} and transformer thermal parameters from~\cite{gao-2016} and~\cite{iec-2018}.

\subsection{Model Validation and Robustness}
\begin{figure}[t]
    \centering
    \input{fig_clusters.tex}
    \caption{Clustering results using PCA weighted scaled features. Axes show the magnitude of the first two principal components.}
    \label{fig:clustering_results}
    \vspace{-0.5cm}
\end{figure}
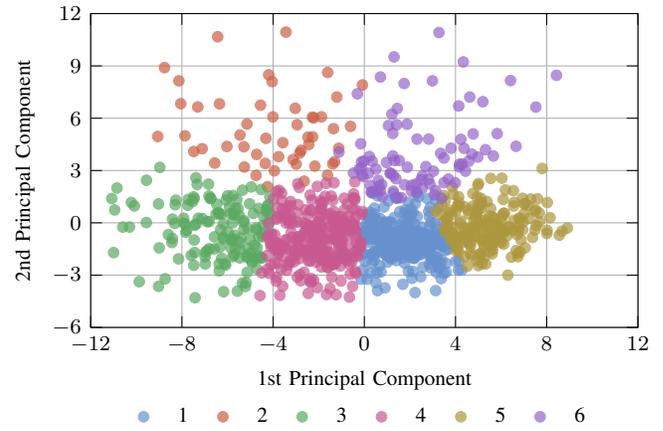

The clustering methods discussed in Section~\ref{sec:methodology} were evaluated to ensure transformer diversity was captured without over-fragmentation. The scaled-feature approach with PCA weights (Fig.~\ref{fig:clustering_results}) achieved the best holdout percentile calibration and was used for subsequent analysis.

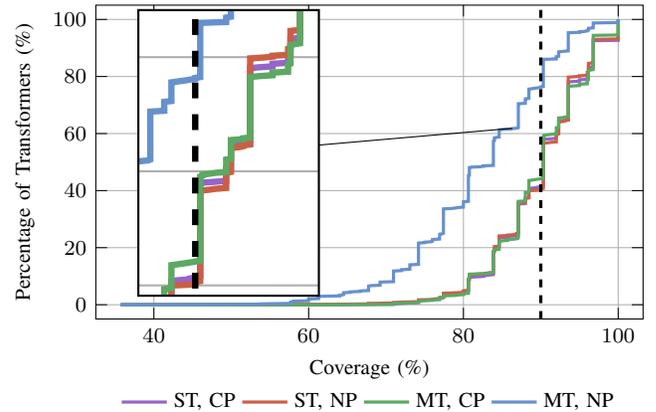
\begin{figure}[t]
    \centering
    \input{fig_Noisy_Temperature_Analysis.tex}
    \caption{Coverage of true scale factor by predicted percentiles per transformer (\(n=644\)) for different training and test temperature scenarios. ST and MT represent Single Temperature and Multiple Temperature training respectively, while CP and NP represent Clean and Noisy temperature Predictions respectively. The vertical axis gives the percentage of transformers where coverage of the 90\% prediction interval is less than \textit{Coverage}. An inset shows the similarity of the coverage distribution for the single temperature training approach under clean and noisy temperature forecasts as well as the multiple temperature training approach under clean forecasts.}
    \label{fig:multi_vs_single_temp_coverage}
    \vspace{-0.5cm}
\end{figure}

Figure~\ref{fig:multi_vs_single_temp_coverage} shows model calibration results for the two training and forecasting approaches as cumulative coverage distributions for the 90\% prediction interval.
A perfectly calibrated model would transition sharply from 0\% to 100\% of transformers at 90\% coverage (see Section~\ref{sss:quantifying_performance}); points to the left indicate over-confident predictions, and those to the right under-confident ones.
Some deviation from ideal calibration is expected given the dataset's limited time span.

The model trained on a single temperature value achieved 90\% mean coverage across all transformers, even in the presence of noisy temperature forecasts. Training with multiple perturbed temperature values led to significantly poorer calibration, 82\%, when the noisy forecasts were used, suggesting the model overfit to artificial variability and failed to learn the true mapping between temperature and scale factor. This highlights the importance of site-specific temperature accuracy rather than synthetic data augmentation.

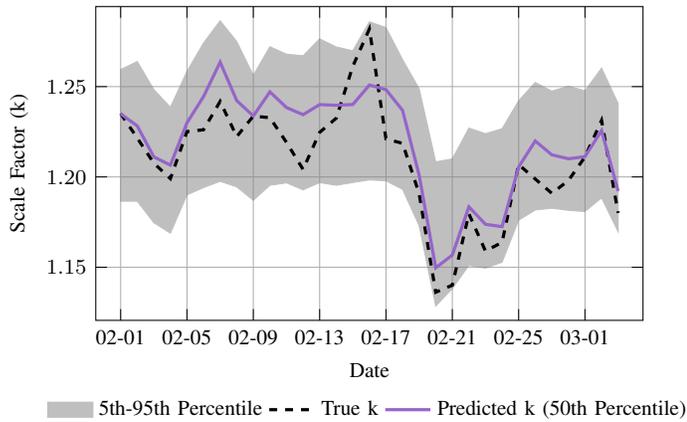
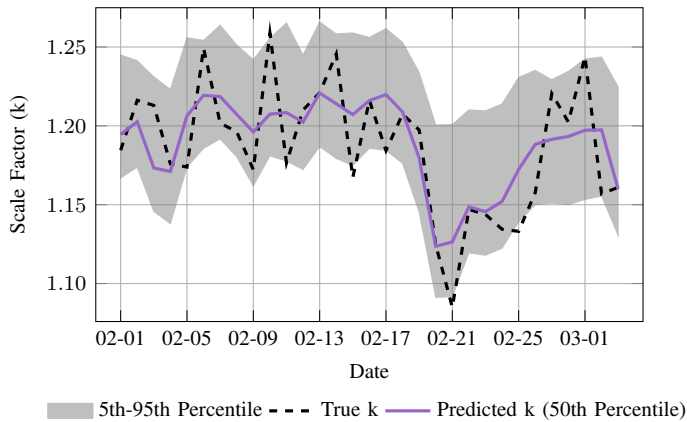
\begin{figure}[t]
\centering
\subfloat[]{%
  \makebox[\columnwidth][c]{\input{fig_Highest_Coverage_Transformer.tex}}%
  \label{fig:best_coverage}
}

\subfloat[]{%
  \makebox[\columnwidth][c]{\input{fig_Lowest_Coverage_Transformer.tex}}%
  \label{fig:worst_coverage}
}
\caption{Scale factor \(k\) probabilistic predictions for the transformers with the best (a) and worst (b) coverage (across the full sample of \(n=644\) transformers) during the holdout period.}
\label{fig:coverage_time_series}
\end{figure}

The predicted scale factors are compared to the true scale factors in the transformers with the best and worst coverage in Figs.~\ref{fig:best_coverage} and~\ref{fig:worst_coverage} respectively.
From these figures, we can see that in the worst case scenario from our sample, the optimal scale factor falls outside the 90\% confidence interval on six occasions, leading to a coverage of 74\%. Despite this, using the 5th percentile of the predicted scale factor would only lead to a maximum hotspot temperature of 140.8\(\degree \)C, suggesting that even in this worst case scenario, there is limited practical risk. In the best case scenario, the coverage is 100\% suggesting the model should actually be more confident in its predictions, leading to an even greater capacity increase.

A sensitivity analysis was performed to quantify the impact of temperature forecast errors on the scale factor prediction. Across all transformers, the mean sensitivity of the scale factor to ambient temperature ranged from {-0.0085\(\degree\)C\(^{-1}\)} to {-0.015\(\degree\)C\(^{-1}\)} for temperatures between -20\(\degree\)C and 40\(\degree\)C. This would lead to a maximum error in the scale factor of between 0.017 and 0.03 based on temperature forecast deviations within 2\(\degree\)C, which 92.5\% of UK forecasts achieve~\cite{metoffice-2025}. Operators can mitigate this by selecting lower percentiles, effectively absorbing forecast error into the risk margin.

Overall, the selected model demonstrates robust performance and well-calibrated outputs suitable for applying DTR under realistic forecast uncertainty.

\subsection{Comparison with Load-Based Multi-Stage Methods}
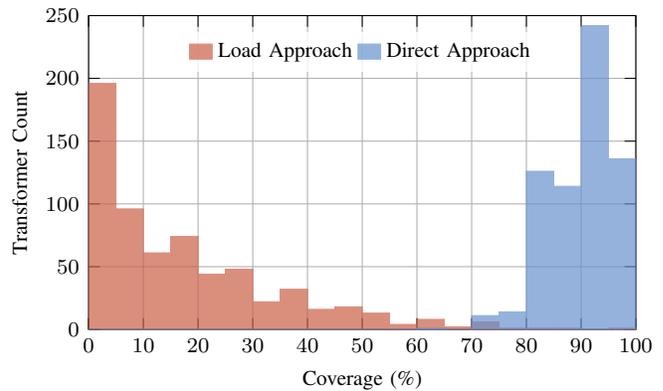
\begin{figure}[t]
\centering
\input{fig_load_direct_histogram.tex}
\caption{Comparison of coverage of true scale factor by multi-stage and direct approaches for each transformer. Nominal coverage is 90\% with mean coverages of the two approaches 17\% and 90\% respectively.}
\label{fig:load_vs_sf_coverage}
\end{figure}

To benchmark the proposed direct approach against the load-based multi-stage method (Section \ref{sec:methodology}), additional models were trained to predict the per-phase load during peak and off-peak periods and subsequent predicted scale factor. As shown in Fig.~\ref{fig:load_vs_sf_coverage} and previously, the direct scale factor prediction achieves 90\% mean coverage across the transformer population, whereas load based approaches achieve only 17\% coverage. This poor performance is due to the compounding of errors from multiple load predictions and the inability to capture phase interactions when calculating the scale factor from independent phase load forecasts.

The load prediction method is at a disadvantage due to the assumptions made when calculating the scale factor from the predicted loads, and future works could explore more sophisticated approaches that could yield improved performance. Nevertheless, the benefits of simplicity and effectiveness of the proposed direct approach are unambiguous versus the load-based multi-stage approach.

\subsection{Risk and Capacity Trade-off}
\begin{figure}[t]
    \centering
    \input{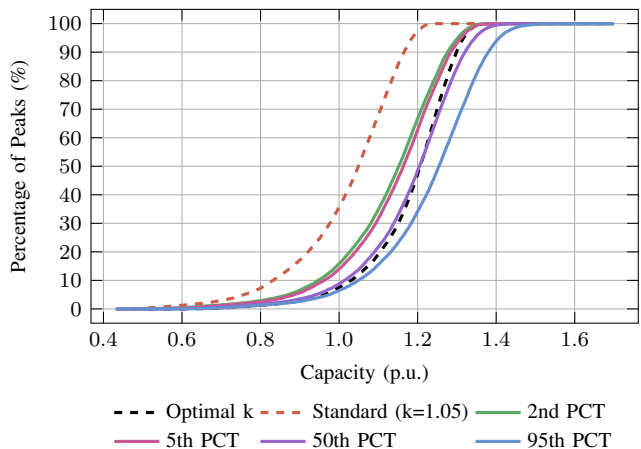}
    \caption{Maximum daily capacity of transformers using various percentiles of predicted scale factor compared to standard scale factor of 1.05. Vertical axis shows the percentage of peak periods with a capacity below \textit{Capacity}.}
    \label{fig:capacity_cdf}
    \vspace{-0.5cm}
\end{figure}

\begin{figure}[t]
    \centering
    \input{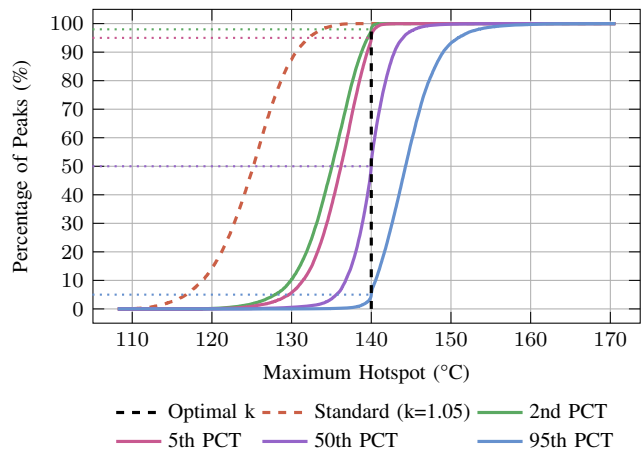}
    \caption{Maximum daily hotspot temperature of transformers using various percentiles of predicted scale factor compared to standard scale factor of 1.05. Vertical axis shows the percentage of peak periods with a hotspot temperature below \textit{Maximum Hotspot}.}
    \label{fig:hotspot_cdf}
    \vspace{-0.5cm}
\end{figure}

The key benefit of the proposed approach is the control of the trade-off between transformer capacity and overheating risk through the chosen prediction percentile. Fig.~\ref{fig:capacity_cdf} shows the cumulative distribution of daily maximum transformer capacity for several percentile-based methods, and Fig.~\ref{fig:hotspot_cdf} shows the corresponding hotspot temperatures.

As seen in Table~\ref{tab:summary_table}, even conservative policies yield significant capacity gains. The 2nd percentile setting increases mean capacity by over 10\% compared to a fixed scale factor of 1.05 while the risk of hotspot exceedance remains at 2\%. This significant increase in capacity would allow DNOs to take advantage of DTR techniques during peak loading periods, and the consistent mapping between percentile and risk can be seen in Fig~\ref{fig:hotspot_cdf}, validating the probabilistic calibration.
\begin{table}[t]
    \centering
    \caption{Mean maximum capacity and hotspot temperature of transformers using predicted and standard scale factor}
    \begin{tabularx}{\columnwidth}{lYYY}
        \toprule
        \textbf{Percentile} & \textbf{Mean Capacity (S.D) (p.u.)} & \textbf{Mean Hotspot Temp. (°C)} & \textbf{\% Hotspots \textgreater140\(\degree\)C} \\
        \midrule
        Fixed (1.05)          & 1.020 (0.135)                               & 124.9                              & 0.0\%                             \\
        2\%              & 1.127 (0.137)                               & 134.6                              & 2.0\%                             \\
        5\%              & 1.138 (0.135)                               & 135.7                              & 4.8\%                            \\
        50\%              & 1.181 (0.129)                               & 139.9                              & 49.2\%                            \\
        95\%              & 1.230 (0.137)                               & 144.7                              & 94.8\%                            \\
        \bottomrule
    \end{tabularx}
    \label{tab:summary_table}
\end{table}
This relationship between prediction percentile and overheating probability lets DNOs manage DTR risk intuitively by choosing a percentile that matches their tolerance to gaining extra capacity with increased hotspot risk.

%% file: fig_clusters.tex
% This file was created with matplot2tikz v0.3.2.
\begin{tikzpicture}

\definecolor{crimson2143940}{RGB}{214,39,40}
\definecolor{darkgray176}{RGB}{176,176,176}
\definecolor{darkorange25512714}{RGB}{255,127,14}
\definecolor{forestgreen4416044}{RGB}{44,160,44}
\definecolor{lightgray204}{RGB}{204,204,204}
\definecolor{mediumpurple148103189}{RGB}{148,103,189}
\definecolor{sienna1408675}{RGB}{140,86,75}
\definecolor{steelblue31119180}{RGB}{31,119,180}

\begin{axis}[
legend cell align={left},
label style={font=\footnotesize},
legend columns=6,
legend style={fill opacity=0.8, draw =none, text opacity=1, at={(0.5,-0.22)}, anchor=north, column sep=2ex, font=\footnotesize},
tick align=outside,
tick pos=left,
x grid style={darkgray176},
xmajorgrids,
xmin=-12, xmax=12,
xtick style={color=black},
y grid style={darkgray176},
ymin=-6, ymax=12.,
ytick style={color=black},
ymajorgrids,
width=\columnwidth,
height=0.65\columnwidth,
xtick={-12,-8,-4,0,4,8,12},
ytick={-6,-3,0,3,6,9,12},
xlabel={1st Principal Component},
ylabel={2nd Principal Component},
tick align=inside,
tick label style={font=\footnotesize},
tick pos=both,
]
\addplot [draw=pscc_blue, fill=pscc_blue, mark=*, only marks, opacity=0.7]
table{%
x  y
0.404249337272019 -1.85496614373772
1.89519459712656 -1.29553790787245
1.82788530163287 -1.20179808794963
2.82692220359336 -1.82935018314567
2.14029154195775 -2.02713571285513
0.0400114829168472 -2.20648969022444
0.232953633459533 -0.565440976398308
0.0673132128376794 -0.422249804407599
1.44869098164109 -1.815176359316
0.409754327359467 0.303344528077089
2.98126420670414 -2.04897675136832
0.741863974484539 -1.69899637322757
1.51579957526155 -1.81167892196324
2.23409149438552 1.10044777324028
1.17825454438189 -1.07467870417712
3.13524149100249 -2.20363957483641
1.37324033973573 -0.581598311109429
0.844190574249549 -0.574978366480033
2.02892385545855 -1.73780787982116
3.56561673179848 -1.57040657003138
1.06521015217235 -1.86416252468337
2.90315723208614 -1.69484861101233
1.04207724243161 -1.53707990197539
2.48766059323173 -0.833300618949037
2.43691371057605 -0.891728704793238
1.76861802165578 0.142826870720009
2.90233866677409 -0.636132645035572
0.986723942017543 -1.85338465130671
1.88894556519127 -0.0062532862770448
1.39507175833692 -1.90533882475934
0.441667572673631 -2.15010118192836
2.19583498662839 -0.693725944226356
2.19395723043796 -1.98522822098741
2.94063305387001 -0.934086010967703
1.94798866254957 -1.94931959396886
3.1292540098802 -3.88861443735197
1.19244578759825 0.017534703616777
3.44938138231179 -0.868723263333773
0.885237951132531 0.0458853353066537
2.92335485282559 -2.59309233084492
0.451647438281639 -0.32779656965279
3.0958802001405 -1.17841771724776
2.4921143528585 -0.590814665403891
3.11500955549455 1.31540823042004
1.68820085078231 -0.873212944867403
1.76991942038328 -2.31640629755284
3.18955743949991 -0.386772016735633
0.650932627360145 -1.03228101098243
1.27400565331571 1.68818688341623
2.13350157288174 0.601531280309196
2.82330169792681 0.648109797388549
1.77534375118855 -0.367460161775249
2.75872898368941 1.39336671093346
3.4828564858694 -0.745914796078693
2.87944647925517 -0.76260159180996
2.85053259228579 0.357838383821501
1.43845389040799 -1.57157945459261
2.26200971396674 -1.29394975468302
0.703352066704171 -0.818082331764979
1.4619979706152 -1.38026115374647
0.261412403377429 0.168453676688722
1.78160287025679 0.729298493082036
3.5018767689199 -0.788694785428569
3.24995958420593 0.840756578831253
0.0852031794035031 -0.056427141590853
0.426111910253488 -1.69852834203656
2.93116277394179 -0.833246462344632
-0.0623669658838077 -2.70011022532714
0.0925664660317516 -0.224551656227972
2.08042104724422 -0.5928877155496
2.86712452393631 -1.48338638690357
2.35755072714958 -0.471908686128996
2.85485408893817 -0.202131715396283
3.38866619624246 -2.1292117121606
0.128145970563721 -0.791862124969834
0.2483323415332 0.414624423256003
2.05230919738897 0.785601468962682
2.59707368424264 -0.607495438447399
3.42652362303952 -1.295074209296
2.32266053259866 -1.78792504718046
3.31088556191997 -1.01904268824868
0.550023290947947 -2.86758866617421
0.435962827201031 0.844208113821208
2.63192714579525 -1.38232523104712
1.75212729429078 0.553445564782066
2.2473395079179 -0.446923765673355
0.879084243348542 -0.344531253972242
2.88380925758132 0.875332678463022
0.130491972958626 -0.946863480730307
3.3023911172002 -1.24868623151841
3.54270343680758 -0.297203174240725
0.70033887246245 0.237869779623658
1.05573307396057 -0.943026093047782
0.583850875223945 -1.37900208352324
1.74651039225601 -0.493233526609221
2.92803886151381 0.0336318397023236
2.58103738280294 -1.37487727255279
2.1344184718265 -0.862576508744909
2.40363400331696 -1.8069068022983
1.05612780920475 1.26051109083691
1.58948720981124 -1.56540196073548
2.81599616919769 -1.23123286280565
0.947296362586536 0.0293672646843663
1.53511299392094 -0.198189342402056
1.46516705522428 1.44941321914005
2.96403533790076 -0.922330819233619
0.453912153141439 -0.898006715138429
3.72209195290893 -0.987201830232009
3.44166849788997 -1.02056258526682
2.63289298105503 -1.39335088565838
1.89839318965813 -1.16497059039191
2.19271625884794 -1.20104314361662
0.980013778639901 -1.53491750904014
1.03740456008901 -1.3066684998541
2.32004743527002 0.362330611220459
1.39124533165538 -0.597930865602619
1.79945646992631 -0.584086496647414
2.6916547445609 -0.749114837714759
2.8556749999707 -1.78014765123839
0.668384145020611 -0.333362529272273
1.92512030059045 -0.579901685738229
2.16695311973132 -0.558181036689758
1.82044363608376 -1.80656669540879
2.87278440891208 -1.63212850879977
4.13995001499661 -2.16536607302465
3.27284966003311 -2.62375742208894
0.240086127640518 0.594228503181689
2.33340731176928 -0.798574471529301
1.53787626085067 -1.29133207786583
2.51992527921644 -0.245214885547064
2.00240211387418 -1.6054352901242
2.06693729519894 -0.819185001602301
2.47436260221731 -1.82258273266886
0.721076499543529 -1.23545094924493
0.946618061046089 -1.84889872586488
0.331383031853735 -2.30193532957051
3.47133548189803 0.0634379793498603
0.424455850782667 -1.6554178866227
2.85120722908129 -1.10473984465408
0.996638592520228 -0.309427177497406
3.17599327391344 -0.631850416058708
0.732130780631083 -1.15640719258356
-0.1079707906635 -0.5498690022124
2.02245048980523 -1.64035162309841
1.30464214373294 -1.21327203777594
3.08994954249576 -0.788189525362466
3.86921103302453 -1.24938981650705
2.47610603295056 -3.09457521092727
1.30003489047684 -1.94223784422213
-0.141794334860387 -3.23231405885812
2.08626008468556 -3.18447580893186
4.12781967824767 -2.71827442061523
0.263580208783108 -2.88066180200783
1.3089768711211 0.306833377366961
0.161537248099641 -0.771617857027818
0.7322160807742 -2.81112491418699
2.93433076487178 -2.11661655415506
3.23931343172682 -2.51874050009966
1.59766582672157 -1.60658744325208
0.410943243652839 -1.87252603823431
0.207844064847323 0.252302222902377
1.4082798148307 -0.84963458322009
2.3455334202182 -1.42798538259667
1.93638766675037 1.17990227775933
1.48271329146539 -0.918951695342536
1.95220288592134 -1.13245643065238
2.12681126599772 -0.663696979083349
1.93548867766913 -2.77740004962066
2.22297189072383 -4.0000064565652
1.10459598539118 -1.98642762699675
0.115909641373835 -3.2339676330244
-0.00989760334056622 -1.75398596412155
1.68818031236528 -3.52384072252488
3.21107026496406 -1.98409580438884
1.04605408370587 -1.79300951650255
0.377698293320657 -0.482755002543088
0.0997471771512186 0.458939051238489
1.770328268298 -0.998498229057156
2.3051375649896 -0.798914484367109
2.04241792898113 0.481186010431344
3.20696981928742 -1.43690586402693
-0.0380475394751609 -1.81280682867911
1.19627711904448 -0.988606563314061
1.66802765885431 -0.8421652060696
1.18702835090086 -0.534799103505382
2.61180209183821 -1.06516130361845
1.62720578722584 -0.773470010074863
3.58559656274876 -2.33306620209826
0.652775367522805 0.0602634219847513
3.33315972553815 -0.618758469914604
1.97583158386811 0.414820002564017
1.46080636232159 -0.550819571817396
2.73496254196743 -1.14636893386069
0.579279300440336 -3.89020936433793
0.941538281500841 -1.61395720120522
2.98467612048129 -2.08199692822341
3.3467278801309 -2.3811174725065
0.982145855406129 -1.08277758364917
0.415376021695413 -0.478091533618205
0.52131525969298 -1.67187211433347
1.40630168061002 -2.1664935024947
0.409530770208285 -0.591212593669562
0.338262157904521 -0.94543039428385
2.07839656494895 0.100541680654382
2.9157807666932 -0.931540568347261
0.378659187704246 -0.76767098072605
1.67644217435683 -2.0173626682393
0.505923079707561 -3.43542859245609
2.57155534886472 -2.5421696926936
0.300195016999917 -1.79995857490782
2.37950407919289 1.49112853701152
2.20175074273194 -0.0843258345668212
0.859123950344171 0.900177725612017
2.29541658007528 1.68127083942417
1.16740400061953 -1.04257932482671
0.357430305958339 -0.122592235858744
1.03409795787771 -1.01384893802814
-0.080084739741777 -0.610383466720072
0.0280171864453909 -0.485833293400951
1.17412843848085 -0.597326021436815
1.49240857179408 -0.446960707301767
0.466900267272715 0.497595571489683
2.40684157709048 -1.47829413764972
0.371017444521578 1.70090194631385
1.93076385315462 -0.172931673992514
0.884822624372594 -0.757797006893371
0.202347029132997 -0.271163762077156
3.4785010508677 -1.08584902670414
0.359566586365792 0.327110694988281
2.67261526878212 -1.51589071862866
1.44250600094039 1.40522735145483
0.481245814720041 -0.590972522222004
3.18609123796575 -2.03181835470282
3.09175853595033 -2.12860933033975
1.67307943049014 -0.0496798458953522
2.17808758864597 -0.554464048718064
0.167442455816973 -0.767852852446863
2.20354879447083 -2.03959884367407
0.584311715265078 1.54561659088022
1.01131874862034 -2.25998489735893
3.48358696686728 -1.36353521880788
1.85679540047117 -1.13464837142717
1.09833143980437 0.0588236846358967
1.38526629869579 -1.40361627694769
1.17189012651177 -2.06847616624742
2.12954781691179 -2.78566596416849
0.668505636547605 -2.46109986994966
1.55579350476236 0.645106343569884
3.07453593899789 -0.0197435701188553
2.27133116688907 -0.731422244580601
3.48396348216847 -2.8987285675022
4.0018195396399 -2.20498253359724
3.59110269141717 -2.82731487790186
1.95050470207047 -0.89026465267276
2.92939568208844 -1.3512132409974
1.69483385782676 -1.03477355850416
3.66146443232321 -1.51773400665546
0.158028480166873 -1.11397238785282
4.27513722162804 -2.71054986936891
3.15677508094573 -2.84344839939552
2.84733979702372 -2.36927790015622
0.904400013105325 -3.7287714567123
3.37202967362012 -3.19952932349959
};
\addlegendentry{1}
\addplot [draw=pscc_orange, fill=pscc_orange, mark=*, only marks, opacity=0.7]
table{%
x  y
-2.21203424342997 6.01897685337863
-1.88185870776549 6.07841761032444
-1.38455947410827 3.38110689989788
-2.27674587204495 6.0605342980185
-1.11869136581613 4.27660971515486
-6.4311319703159 10.6740004814947
-2.98849018336285 3.36766503574949
-2.26639833167713 4.93322601923491
-5.26286602138246 3.20270972655098
-2.63834993189195 2.38862728032316
-3.02935336669508 6.56757061640884
-2.83151086404651 4.00311910106636
-4.55175788200193 6.74696747704523
-7.86495067891625 4.99340495121377
-4.31788194015858 4.84556194567013
-3.12426894768673 3.62795269533154
-3.76040865473551 3.79332339438816
-2.43599230267423 4.49332206546985
-8.05063497729191 6.83099646983494
-1.30092025285541 3.50645113439036
-2.47008917241649 4.9116977853041
-7.30358681359588 6.64667979328824
-6.02669032333122 4.37965381524904
-9.05871534051313 4.95154870496358
-2.16885149936409 3.2544538936625
-6.561252144791 3.42775247398977
-4.04280789988613 8.10031122811552
-7.48128349023195 4.0918610174231
-2.94577538619307 5.62799969739624
-5.28042908539199 4.36546565903925
-0.082391192970146 7.90834911181665
-8.76384742722918 8.90184189715625
-4.19350568492881 8.49043867783389
-1.60978481112935 8.6236175731148
-4.24617286178824 2.07242495273366
-4.34483982004222 3.40494782243409
-0.60460271418445 5.53248053429382
-3.43698102231478 10.9387309161176
-4.74849439542152 2.71971093543523
-2.77032462962837 4.15226311165946
-4.00997683472064 6.07865176819902
-1.35083021816876 5.39934062685978
-5.45304614483048 5.03624921728271
-6.35120895186358 6.82290900055259
-1.20568533834744 7.21269491824936
-7.11230973679831 4.24954170394989
-4.72043301789187 3.92154909073341
-8.13129249883227 8.14572153934278
-1.61583529727433 2.72733445011528
-5.14688897610429 5.66742458406231
-3.57268352939547 4.59625879900576
-3.9441074359538 2.98249427802654
};
\addlegendentry{2}
\addplot [draw=pscc_green, fill=pscc_green, mark=*, only marks, opacity=0.7]
table{%
x  y
-8.31846769039082 -0.0330466312052045
-4.63337256584021 -0.804653475347817
-4.62412483234983 -1.90063831460431
-4.73672160725312 -2.03431569172055
-4.24507234679604 0.931935993778737
-4.17482239077977 0.000207881254146836
-5.86161614941586 -1.42265035279104
-4.54942960941023 -2.016228651888
-7.84185002521775 -0.0788352541882711
-4.93204552334498 -2.20341076736039
-9.56054634232822 0.0146313187174512
-5.48784035178283 0.410953115638852
-4.52735248064525 1.91190520273397
-8.59637658533012 -0.674422244097281
-6.97489893059694 -2.35680273857245
-6.17190905650352 -3.95057337468946
-10.9946847334698 -1.70959404223795
-5.03353519626831 1.09088328859476
-5.75766085101624 1.56952209327358
-7.11227428197266 0.248307804229976
-5.28219953563509 -0.387054646038685
-5.76909422747471 -0.916058331397613
-5.01346636892292 -1.75762247191326
-7.43738378603922 -1.70244702447598
-5.8766970534996 0.524232119320301
-6.36894140165854 -2.35663707626427
-5.34031470803005 1.87375924976259
-4.13022692170651 0.510283679888527
-10.2981419491607 -0.242093782143237
-5.09363181352221 -0.345475958439439
-4.74301108353485 -0.40053564163656
-6.3612374268267 0.521356884316933
-4.82415809981621 -1.30286142044325
-7.48742406867911 1.48530850076943
-6.901349956492 -1.60270814009045
-4.62854756300833 0.829434687146116
-5.15812839188872 -1.91123035500839
-4.69554564081915 -1.43161768408339
-6.47467270048258 1.07378939685312
-11.0772542652876 1.39314469826945
-5.78600872885604 0.000221859288212167
-10.8487270156984 1.99618462609467
-6.80121767018379 -1.48587487678228
-4.8323100744262 -2.00396341208518
-5.34866190819266 -3.45878938717354
-4.47313842664688 -0.583393996820586
-6.33712496632517 -1.39904246344053
-4.13202064881562 -0.641273587860471
-4.4812469368962 -2.00919418033757
-4.71086367535019 -2.64970330793467
-6.27065978275257 0.492798712925332
-6.50509152490876 0.508725538312007
-8.73297167132386 -3.20987479098119
-6.97227832080328 -1.69762767526517
-5.80619768831067 -1.02628789053547
-5.56549323832642 -3.91867791844734
-6.11676320122843 -2.34642257389196
-6.0837158991228 -2.41707681030939
-4.45765354646278 -0.795942497249215
-9.02568666654747 1.08303511271162
-7.42338047517627 -4.29760786316862
-5.77386130581085 -1.94251177688778
-5.1904507010333 -3.42854521601503
-6.36391261033753 -2.75719833797661
-8.99492860496921 -1.43750005977812
-5.4025073158799 -0.874913320272215
-9.87160396968292 -3.37451673127337
-10.2185166288753 1.16947169659681
-7.53132117601192 1.21867263033302
-4.7309173660719 -0.29934357275816
-8.26902151841885 1.30016622043826
-4.42630867162604 -0.128690390168233
-6.81051934533409 1.54131180937262
-5.14029934205352 -2.83985119394749
-5.65831679995661 0.832757256905551
-7.38634674208557 2.58341263441438
-8.29213873484267 1.0384308557457
-5.27253761454042 0.124157919598857
-5.31163860452594 0.0447359726611525
-5.7414744662689 0.521455888709026
-6.06293481188838 1.30019899288692
-5.59183003600551 1.34324570855413
-5.69228806544817 0.457722352607441
-7.46299484650088 1.38977340670042
-4.95706290291099 2.06061279959586
-6.16386454998609 -0.945822794541762
-6.80407362574244 0.0746601897560326
-5.90756011348869 0.917902575282198
-7.38298943610883 -0.0783583084197563
-6.40537261140897 0.58297377018353
-6.2575246291046 -2.03370149960321
-5.99110101156166 -0.526219251483083
-6.30310867654209 -0.395098601917335
-7.72428564141974 1.34257547352508
-6.18823823951374 -0.833873474884326
-4.23703871603004 0.548017495128263
-7.4616003441347 -0.985604769478804
-4.86650131523535 0.536605048875375
-6.83243746148989 2.21982402986661
-6.89463514316529 0.0675461954547021
-8.96841383005723 3.18120071299295
-7.92366740193721 0.266453438505617
-7.93237125724155 -0.00661556800089413
-5.10581244175432 -1.46470348740128
-4.30262530565482 0.749985938816577
-8.43594764337364 -1.12933730279559
-7.78600290650521 -0.424311877629915
-5.47080943420344 -1.52575711748032
-6.75531209627925 1.39205879822485
-7.33050038060029 2.20256165716002
-6.72985559034686 1.40911847726244
-5.00366622962477 -0.965879825116529
-4.41425672487409 0.370110261243435
-5.53552060660651 0.821743153381633
-7.59357525027036 -1.86834683305133
-7.83046805249948 -0.546973560680688
-9.55426957330736 2.44463405886004
-5.71499711670499 -1.85144947138397
-4.22464227631961 1.02187007672056
-7.74335808082535 1.00704271860781
-5.47263565422431 -1.09937840475054
-4.92517742930799 -2.27613416599777
-5.21584727324048 -3.64739143236141
-7.47977644765026 -2.43383387365937
-6.98231711093648 -1.9480930331068
-6.5714542641064 -0.800113467422896
-4.14408851788197 1.50594273193518
-8.41468933206216 0.212820801064378
-8.41547355804696 -0.900476340142689
-5.13485274947092 -1.97804215637559
-7.2852811032316 0.259888353738112
-5.99488410318679 1.55089181105318
-6.18093266134907 -0.051704773398239
-7.4405909660993 -0.136922223487047
-5.46673478364406 -2.00999748282201
-10.5951481249603 -0.244498365703153
-5.68878751308887 -0.711449376748498
-6.58055050639291 -1.49684042592026
-10.0849745399104 0.970363036939968
-5.64899515190463 -1.44956788710869
-10.9475868804093 0.741050710656809
-5.83738280855009 -0.943005548787706
-8.15002257805378 -2.06524563463036
-9.02394126635695 -3.64415214879835
};
\addlegendentry{3}
\addplot [draw=pscc_pink, fill=pscc_pink, mark=*, only marks, opacity=0.7]
table{%
x  y
-1.05253365999982 -2.47341668095894
-1.08278739972512 -2.04553824683596
-0.790378471655658 -2.14976415700342
-2.11131308771271 -1.61977896712635
-1.92290302595463 1.25685114336293
-0.934136093780686 0.164138620318855
-3.42657025053429 -1.4705997558483
-1.02116980044038 -1.0059810028196
-1.36166712517321 -0.254937739928068
-0.13871805991374 0.48360634401027
-2.36619361085388 -0.17665472249794
-1.21730523420045 -0.00308974959275177
-3.25626466049993 -2.25803061777039
-1.8846871579996 -0.172101305255529
-0.34805228725788 -0.527006559647956
-1.24664273299016 -2.28188588640683
-0.772706378905419 -1.87385453981251
-0.860123493997748 -0.0439107456007006
-3.65378979840391 -2.65797375567105
-1.61845469415542 -2.55761343632589
-1.46580484752501 -0.74884816235639
-1.65733717165068 -4.10876071824071
-2.29363089601937 1.30595483864286
-1.23484989708413 -4.11866257779983
-3.57887340090483 -3.63800412957468
-3.7745340889412 -1.79594043250167
-1.27633159609095 -1.91314002484975
-3.84973638624041 -3.12068262181619
-3.70504501150615 -4.15327517861129
-2.48905019533242 0.439363925453879
-0.87170809621965 0.738186118013743
-0.419915616888863 2.34483400768798
-1.38797319534138 1.28356661119522
-1.34552275910888 -0.903938390505223
-1.05826327953765 -0.536950979969671
-1.49526230170419 -0.662050051202945
-3.34008859718843 0.145362864816047
-3.80660070926355 0.874434325222765
-1.80832371741371 -0.291248059169047
-1.14195550673188 0.172569644175209
-2.86624603458708 -1.15123673393165
-2.36572053961054 0.731332158424356
-1.05546953185922 0.359282218304761
-1.45100192141561 1.5324754408534
-0.364723518488532 0.368970258917509
-0.662503930145344 -2.36246249049789
-0.868796554975314 -0.407852782489407
-0.205729111021316 -1.32453369251718
-1.89848279698805 -0.312134775698969
-0.957620077095459 -0.943282669434665
-0.0916721325796479 0.652091034538223
-1.16730154426449 -0.884897690502091
-2.01321064747485 0.0747858320539017
-1.16883499814552 -0.232090758414794
-0.974604193992044 -1.76823642065989
-2.93945979252494 -1.3819662465482
-0.166213084363737 -1.42474635035216
-1.24162870184751 0.35054253680041
-1.04516498089586 -1.04580193143431
-2.50509872556465 -1.72742154479009
-0.453745685152076 -0.190970011169707
-1.00564742498513 -0.185518901270858
-0.490928085043257 -0.183310319248922
-2.19413923957571 -1.58347583802112
-0.470370196442596 1.65562607642376
-0.352519154276608 -1.14643525918123
-0.997202328122977 -0.292427367518656
-1.32901948275189 2.29554258813583
-0.595453030660557 0.129350723316896
-1.31887205154342 -1.99583799087212
-1.05346504894025 -1.91566836969803
-4.24990375285068 -2.66473545860816
-0.612776511882586 -2.94485367464003
-0.672943580765254 -0.700014574877878
-2.57208372562951 0.744859663938326
-0.906208256169383 -1.95945301127219
-1.3685793233505 -2.4107517445639
-2.64749766093775 -2.83423410610616
-2.94429643168604 -3.10714830888656
-1.65024894314423 -1.94076205139039
-1.49298568882789 -2.21011032019533
-3.33642563778873 -2.41772878799006
-0.406290348911011 -2.77533087566853
-1.33815996041271 0.656224657330663
-0.514333841771466 -3.64022779636132
-3.86859957943061 -3.34290203271179
-1.50547490581884 -3.76603519343586
-1.9073829328268 0.830263681786907
-2.45916832247227 -4.04719922791236
-0.804012313867453 -2.0714107525601
-1.72271149559757 -1.59981215731214
-0.710853605378232 0.374955694657502
-0.3765410180131 -0.83453754544307
-2.65708062777201 -0.595917815760659
-2.11763125904121 -2.23901747731889
-4.05477315346398 -0.631920980567868
-1.02402228538844 -3.33928351101278
-3.05746998898312 -1.89902912280142
-2.78837872007565 -3.32507192442353
-2.05971742599489 -3.71658696178547
-2.06305126155012 -1.98512939184632
-0.377524608842811 -1.87987339979884
-1.5571592549325 -2.16663614444321
-1.4472295858225 -1.72092356066423
-1.34785899629114 0.0660534296610049
-3.84752123887552 -1.46602285544023
-1.42939457995868 -3.11616765511107
-1.10020085053668 -1.78489085949853
-0.396647820842374 -0.105640476576606
-0.469597776556793 -0.108807676532909
-2.22493750229869 -3.28895794413603
-0.653695014648985 -4.25768470336395
-0.158365663015526 -1.11755584067265
-0.484713719213999 -0.274540524807374
-0.286611977608717 -1.92029447599807
-3.5691369540546 0.37375192476734
-2.85967041153769 -0.970570780806532
-4.08817983803346 -2.40402078893372
-0.495020275076454 1.42383096880765
-0.155215685478977 -1.15581447616887
-3.20386185956864 -0.54517703813635
-1.88134851094055 -1.46004794834689
-1.36091713014659 -0.914909526131392
-0.991788400558189 -1.18861067170245
-2.49295725002106 -1.87959041453199
-3.36389276084527 -0.96988904909947
-0.853444065915452 -1.73128048845446
-2.21015933203517 -1.66778421716824
-2.25905045213293 -0.442496531790146
-2.30127993866224 0.69619237968026
-3.06892427649115 -1.60498354866831
-1.35705644003797 0.150771282174188
-2.11577195478093 -0.760455551333098
-0.352023735442713 -1.71592795548148
-2.53987204259721 -1.68220066278997
-3.07953311700996 -2.2320132421033
-1.98568981446595 -0.733122971705304
-0.363098211525721 -2.89279553444137
-3.23434071574893 0.39099641287009
-4.03255126993936 -1.29648706676526
-3.40765151949331 -2.72685494395527
-1.089023353735 -1.11648559778545
-2.85285713039604 -2.05610082030827
-1.62631598769098 -1.20634924307899
-3.12054643206524 -1.96243839568589
-3.43371552374941 0.790788496165407
-3.22647000961904 -1.55747145261734
-0.250615814067707 -1.43527002690066
-1.63400637979217 -2.25916902130396
-0.603502396865262 -0.429869089449475
-3.27824920479571 -2.20648128565849
-2.82253247626853 -2.97009007286268
-2.34757034315051 -3.41978992221135
-3.86077566512677 0.201865727904335
-2.2375708228452 -1.17852289841934
-1.59976537035162 -1.67993360430682
-1.72487943315906 1.51953029465533
-1.67846536904404 2.26335187973907
-1.92626124696587 -1.17669349037584
-3.92827245767093 -0.681277816126119
-1.94217606940109 0.272212632795872
-3.37361124791245 0.273171945710615
-2.90607783985778 1.04031292550828
-1.86679776212158 1.13689960159323
-3.73444039163997 1.86410011723397
-3.66799325146375 0.40032785073204
-2.49693117938765 -0.85999198680146
-3.80229952202306 -1.19875872316645
-2.5653804681683 -0.06498246011526
-3.82751841960707 0.333302243719819
-3.67707264062978 -1.2748236037931
-3.43306756206 -0.0131684684946718
-3.83687506402252 -1.1432362057091
-2.44693205392293 -2.19507189106084
-1.84772996617552 0.80894001832884
-0.791220208291492 -0.15617882651765
-0.690043919470797 -0.148088901609006
-0.418635404464947 1.32784401954199
-3.62255257386573 1.2729383399342
-3.02383565914281 -1.00348668532312
-2.06294716453239 2.08320588160033
-0.436038653235011 -1.69931289863303
-2.18021679308944 -0.900724472059437
-3.23235440201286 1.22273006666889
-3.62796705252354 1.69974630653231
-1.4960182868783 0.527525036548834
-2.57028266758106 -0.416741018180647
-3.22003961075128 0.994181757427801
-2.87715108046315 -1.7624450961401
-2.37141564183921 -1.20492634685513
-2.95824375224297 0.567342807996009
-3.59274262797492 -1.91874018251036
-1.92031267716849 0.812733321659722
-3.20415869411969 0.162635878833363
-0.668243379787092 -0.456306650855211
-2.94089113959162 0.0410567444974377
-2.6770918072809 -1.58676981884768
-1.91789090869379 -2.0375763875174
-2.90799325527544 -0.804946017958511
-2.07677315727094 -1.23971983137435
-1.90033169152285 -0.804061205161278
-3.77979916472128 -1.67793290526231
-2.05313013510576 -0.381869215263982
-2.24122735390315 -0.143177890719729
-3.16211331002616 -0.830288045198398
-0.393067876273854 0.745748331359708
-2.51447448836469 -0.736803514171991
-3.14820176288902 -0.0356640637864127
-2.91774684715917 2.25172195805704
-2.85094514844446 -0.807134290918062
-0.995738523283247 -1.37152511126631
-3.97963310995385 1.37559235877236
-0.69405075664679 -1.33825999266276
-2.44704609058231 -0.215703767169478
-3.70156958599439 -1.28304164635883
-1.5951429137299 0.431916015424557
-4.2433483334764 -0.809972917385761
-2.85149630462904 2.02574920531574
-0.417797382021702 1.40564470985379
-3.59370779774121 0.794777641595508
-3.33093739074971 0.934826302976154
-1.16842681418838 -0.836421676764374
-3.62410909410299 -1.34147473106235
-4.20701459647294 -0.0347245801801694
-3.83470734906262 1.79914598808834
-4.12300123011886 -1.15410351590293
-2.83241459808981 -1.41744447519648
-4.20933082833384 -0.709931867434169
-3.77288645487921 -2.3117545017225
-2.68883926721023 0.0199738371190179
-2.23177587954767 -0.988821954371424
-2.4582199519302 -1.56663336625158
-1.76545379174336 -1.76048538754939
-0.508011263409619 -1.97249941413908
-1.59934388239484 -2.55671591849402
-1.69151265968705 0.235048123838957
-0.512327203000506 0.80592818107382
-3.47376932130198 -0.98040661762187
-1.8982598030741 -0.667395046877764
-3.49550908978539 -1.35317710750088
-3.46666028568183 0.529283357082686
-2.3758865272169 1.59116879452244
-0.883315085055248 1.41243437734896
-2.33262190763826 -3.19750634525052
-1.20077962603064 -0.944768173803062
-1.29701732853045 -0.37596210754378
-1.7832526070659 -0.935037852151007
-1.81036485661264 1.17409000824452
-4.17059303334743 -2.90064300438037
-0.743718942009848 1.07943159620858
-2.03041879507086 0.355169134971802
-1.11329638384088 -1.34497476584483
-0.963724046317234 -1.75843477523426
-2.38274546698794 -2.18217614862521
-2.51339199836335 -0.480283735164152
-4.09476017603342 -1.36166702046429
-1.44010154528938 -2.10250267648788
-1.32674854564878 -2.36848134588099
-1.12481324020532 -1.22245442324927
-2.37939611776304 -3.36305231443934
-3.93328338066497 -2.29732864218226
-3.90957124134808 -1.53604816601856
-4.26804041839909 -1.88100943447179
-3.73955050546271 -1.52571151213407
-3.28615527572368 -0.0110968943281835
-1.88838977023592 -2.52803077782433
-4.49043980743311 -2.86970345419427
-4.57345616253606 -4.17191143608807
-2.51762508812746 -1.54177002770592
-1.48161787869345 -0.163141071271031
-1.70396891515669 -0.96722382652472
-0.550553541201005 -2.23299054209547
-1.80640423058739 -2.17467869202184
};
\addlegendentry{4}
\addplot [draw=pscc_gold, fill=pscc_gold, mark=*, only marks, opacity=0.7]
table{%
x  y
3.85704050057644 2.06792819087925
3.50051315706646 -0.296188632675715
5.23552824138515 -0.469384065393739
4.38160990693542 -0.305814608430793
8.90367136657518 -0.296091432419773
5.50071226368345 -1.66134674520258
4.02656490760099 -1.59853626174442
4.76410834767478 1.68408672608134
6.74636430100419 -1.75191221421837
6.08058788390842 -0.597307400083516
5.78120550615443 -0.0232013683648311
4.5774781399562 0.516709751434218
4.13914563132139 -1.6312650996866
4.38886309702309 -1.93760474222651
4.97856751507563 2.15135324552229
4.06227719218706 -1.03518239969374
3.54701388920631 -0.57477885929026
5.41630105433447 -0.850287811228651
3.73595161937934 0.576648116443312
6.75028698343069 -0.434307276176307
3.74751593719356 0.719151232538444
5.5369359460273 0.823041039902741
5.38291937392439 1.2179695325059
3.32792184489427 0.4865578886426
4.06550575601789 -1.45715997690361
5.55590982392756 0.00734902651010394
4.71180439920248 -1.88995515979575
5.42183020153643 -1.33756490580131
6.73894123083926 -1.93620757170799
5.57483054305326 -1.37501664674252
4.84588670449311 -0.218577074009594
8.14567269095247 0.296477569420097
4.9486069388855 1.49410465986165
4.60590110368118 -0.313666990868202
3.89344997495236 1.64068604739541
5.12123144442683 0.126770345641182
8.01539781931566 -0.432730413035906
4.77206604858706 0.958926430699755
3.92291934254462 -0.296857406218536
3.53569644978532 1.25451787750993
4.02248910017985 0.676276756732813
5.06267536596878 1.09553702393994
6.35244955933615 -0.231819428506436
5.60983118005222 -0.123981801681993
7.47942970940155 0.674267679779415
6.55075095926989 0.138433320901677
7.48666390090874 -0.405755462223769
4.09326712083581 1.00017875309716
5.27700360750577 -1.35179017066164
6.7819464949353 -0.211013242824519
5.89275707559685 -0.595515770332535
6.01474219374945 -0.477524785052699
4.23443093253486 -1.03147270114765
4.53993588290061 -0.544496980631427
4.39853241487456 0.921173704780892
5.54463130050388 2.54777021165182
6.77846808834295 0.254282501497535
5.00303864281624 -2.45214709827614
5.15192991226705 1.13678562721112
5.1134787218632 1.39989611198732
7.67205535248798 0.559212651202469
4.98131187057821 0.973739765300813
6.09088011965765 1.34039909349984
7.54455790253479 -0.162931469339686
5.86162615245849 -0.169941220984415
3.67276255985385 -0.0257506101178047
5.21516757461174 -0.181413408981129
6.29083960815479 -0.313801993497026
4.05258025965934 -0.762174427752301
5.44556531730027 -1.0763417879356
5.13370422833275 -0.877221483055967
5.88811061462509 -0.249056805179849
6.86907328465199 -0.974455865727651
4.91821250502759 0.563773161507299
4.18081817662036 -0.00918651909454948
4.8873442826405 0.386623214318645
6.09557320169706 0.541448379909428
4.88869586988589 -0.763473640967961
4.41214892457899 -0.365511364180731
3.90513822851561 -1.02004481882306
5.61230405919106 0.339089673602527
6.00639127947048 -1.13068984799267
4.60016565767928 -1.18780101440105
8.02383915550579 1.16606261627674
3.80618102618171 -1.07042724631852
6.93585453279563 1.40878063500873
6.18798427611602 -0.465281271974552
7.03319417651557 1.0879324475037
5.30625960382398 -0.0810012558038602
8.54883548633953 -0.70730784152482
7.59354579798045 1.38095474173849
5.11441167947752 1.10396501967388
7.18611516667414 0.990886906789708
4.38599770956491 -0.833362077383577
6.69974412060649 -0.673016102149511
6.94677599102933 -0.395186056730597
4.14424538947103 0.956930283385564
5.42615037778934 -0.828379130354024
6.18821092088487 1.88900354666839
6.54502095432271 0.964300353198175
3.66146965833827 -0.254207864273125
3.92316165665634 -0.748323585389407
5.47239862201238 -0.621719137653028
5.08239566371549 -1.77522747338062
4.90299041411237 0.575885107003866
5.8432005052109 -0.375912462058661
3.84248612748596 0.472976407687831
3.79397887890249 -0.586655157797562
6.05273952832048 0.385461398895746
5.54709325006679 -0.694673585250651
5.17207154993804 -0.949182080910543
6.03493506915266 2.56202031819487
5.60296268271062 -1.30529844790664
6.02165611461356 0.0451843565707757
3.80153728292804 -0.762112367586786
3.58940462469361 0.281973406414835
5.60433812707329 -1.78114453810612
5.009619109 2.25358911877686
5.16881575750466 -2.33487593764034
5.32416195388628 -2.35959748824461
4.14995359481041 1.17239630402181
4.60780467489072 -0.858193013765559
7.68593502448952 0.0722711913502467
6.53536453194779 -0.439818460545868
5.08562534746395 0.784957431723157
4.89948247874231 0.204530479105268
5.49043449389145 0.38680029466157
4.13645774835655 1.49407940841859
6.1522151512442 0.544094960353672
6.24969643166436 -0.749709463698464
6.11545946911469 -1.22565167269997
8.67220531554897 -0.4763658734497
6.45156177972056 -1.97185988655473
6.8298212859241 0.260541007980807
4.5195727563467 -2.48602570031678
5.53787300175839 -0.284511376760443
7.29750345684519 1.36523229800979
7.37825348331378 -0.838926418734451
6.05169049083458 -0.53342714782012
5.7103899793981 0.267008257842986
7.00408134389948 0.530392598272093
7.21332344452819 0.247447633138017
4.30481112138926 -0.775569812417181
4.22441375235693 0.102570291044748
6.53821944022753 1.84640814586298
4.65684480932737 -1.33955175545042
4.2114907948712 -0.841900990256595
8.00522903839926 -0.609432390621228
4.84462692158219 -0.42925476933816
6.65990938992769 -1.5898145166389
7.79904746760931 3.11480756528434
4.72547595379485 -0.353609896023405
5.57387876196528 -1.02122897407747
3.86706425786749 -0.556252964499373
4.31775254075453 -1.10983434612394
5.08221762621515 -1.90280276111491
5.80282573943205 -0.848684314091232
7.05548409209286 -1.42923626724435
4.65200253314059 -1.7441022712075
5.84195105948388 -0.481193670795021
5.64672248474568 -1.02699842617232
5.07237927028704 -1.7803081373004
6.83402314086557 -1.66971090822688
5.63067125434643 -1.55743833283764
4.66680137815203 -1.10440336919968
4.77009807568628 -0.739656686190524
3.61736364337558 0.211105936515941
4.56261048552729 0.73226138675857
5.96379755725193 -1.02931102082088
5.67312986178531 -1.05491180754793
4.80223691324953 0.130197229203412
4.76411905772964 -0.307856392962049
7.02285959420565 1.08325878632423
3.13490869648418 0.669200750295375
5.07166069719733 -0.328573087142079
5.12288153887535 -1.01192034338367
5.0047557907136 0.789503300671909
6.56733688218461 1.17403162750951
3.39286086532299 1.02988394836864
5.31112313185688 0.164429696312449
6.96165409099598 0.0393581949397523
4.93515158258789 -2.16142820831949
5.46965638626954 -1.0268458040647
3.48615301944162 1.56564836161973
7.5191375254954 0.742426309335668
5.23425570091325 -2.00331878088765
6.29491876577061 -2.99821645830163
5.24687351732022 -1.2826828581972
};
\addlegendentry{5}
\addplot [draw=pscc_purple, fill=pscc_purple, mark=*, only marks, opacity=0.7]
table{%
x  y
3.68940185427324 2.51100364046121
4.4381930122615 2.96387029413903
4.58798103955232 3.54260525126048
1.38090309173465 1.43360766638778
0.299971436728997 1.75106375619919
4.16425810768286 3.21067666977304
1.88452445112079 3.19324783349263
4.23679173711008 4.78794190589028
2.49234464141577 2.48117383670335
2.52818089646875 1.97957498074709
1.15901271198958 2.8424500548052
1.24701786426437 5.12718859844029
4.28616082916801 3.37360549477447
6.65512491106784 4.38814641327037
-0.47744984102872 2.85038641121497
0.341985167276249 3.53570864840458
2.30004792272903 1.74904779491627
2.31172218794704 4.80095656308206
5.19630940694462 6.94496467525967
3.98618652869619 2.3004116357
4.13847113528499 6.70533841714725
-0.218446100358318 3.08630305245237
3.49585270934106 2.30534995586066
0.509379752359994 3.0888493067542
1.74510486845777 7.99238455295856
-0.101272137452987 2.74675963547785
5.31600641501625 4.19178347775425
1.8724324949691 5.65935635684079
3.49199032080406 3.71730339838879
0.221151010327472 3.80481719794265
2.80377164550641 2.49223959932725
0.334080730689767 1.7813575256048
2.87054070088009 2.22842868291158
0.706969638425976 8.3628796526972
1.20839079499292 6.23091932739034
4.9819772717872 3.77276253907603
-0.0222153355599009 3.54591561991103
2.54323986933097 2.69177835905143
0.54873214294849 1.81383268061184
3.80439174467347 4.27766975496431
2.82060755028167 4.80322791580384
5.82725765689817 5.11613147894487
1.41964777851756 5.626760610942
3.39252715609446 1.41011117938879
3.12663799049665 1.85733554969358
1.76635161023856 2.18505319921739
1.05824376140352 5.57906110653081
8.42730128066453 8.46039732003025
2.07418141292642 1.82015149385777
3.34712630215447 3.98333773675576
1.13102279374745 2.83801795066812
0.259018797743569 2.14832165571982
5.65961771230534 3.83599776413108
1.24841095241569 3.83847191476944
4.71742910884393 3.94651001059657
1.30394777728922 9.51697873931347
0.695995538993983 1.83520447085702
3.07255618052826 1.60803344092081
1.52911038370568 3.57553978282616
3.02956598874299 2.95879049506062
4.37133342006506 4.27754704880978
2.69831585348208 3.22494586184764
-0.0515688994106649 2.65332516682783
0.788066168919225 1.50583528182714
-0.155733608870719 4.53388547998728
0.64765045285475 2.09307816476068
1.27468386268933 2.71512900811549
1.38472773446914 2.53004777064763
-0.114616813853091 3.18048239361205
2.07842719195686 2.14761612578692
0.971123365527953 3.46211821617577
-0.125826722323513 3.49005815610385
0.614013546685323 2.4376604113905
1.3218635697662 3.57582592192257
-0.305421150899406 7.40219487726014
4.34148781612343 9.22877619668589
4.77179768922801 5.09908471796388
1.27219147463448 1.81274627764544
0.18537748574445 2.84095614750994
1.44479920020683 6.55937046835952
0.762751457755259 2.19436740457329
2.98723570051019 8.14700682636896
4.17469403137124 3.36384947454396
1.62079236933306 1.43891221485657
3.27196804726892 10.915657648386
6.41196005882136 8.16030054183586
7.52347395747602 6.63844727794625
-1.11388183546169 4.05833227763727
2.27213463448104 2.23614845667738
4.6225107386513 7.21478569081474
};
\addlegendentry{6}
\end{axis}

\end{tikzpicture}

%% file: fig_Noisy_Temperature_Analysis.tex
% This file was created with matplot2tikz v0.3.2.
\begin{tikzpicture}
  [spy using outlines={%
      rectangle,         % shape of the zoom area
      magnification=2,   % zoom factor
      height=3.8cm,
      width=2.4cm,          % size of the inset
      connect spies,     % draw line from zoomed area to inset
      every spy on node/.style={thick}, % line style
      every spy in node/.style={draw, thick} % border style of inset
    }
  ]

  \definecolor{darkgray176}{RGB}{176,176,176}
  \definecolor{lightgray204}{RGB}{204,204,204}
  \node at (0,0) {
    \begin{tikzpicture}
    \begin{axis}[
        set layers=standard,
        legend cell align={left},
        label style={font=\footnotesize},
        legend columns=4,
        legend style={
          fill =none,
          text opacity=1,
          at={(0.5,-0.2)},
          anchor=north,
          draw=none,
          font=\footnotesize
        },
        tick align=outside,
        tick pos=left,
        x grid style={darkgray176},
        xlabel={Coverage (\%)},
        xmajorgrids,
        xmin=0.325, xmax=1.03214285714286,
        xtick style={color=black},
        y grid style={darkgray176},
        ylabel={Percentage of Transformers (\%)},
        ymajorgrids,
        ymin=-0.05, ymax=1.05,
        ytick style={color=black},
        tick align=inside,
        tick label style={font=\footnotesize},
        tick pos=both,
        width=\columnwidth,
        height=0.65\columnwidth,
        ytick={0,0.2,0.4,0.6,0.8,1},
        yticklabels={
          \(\displaystyle {0}\),
          \(\displaystyle {20}\),
          \(\displaystyle {40}\),
          \(\displaystyle {60}\),
          \(\displaystyle {80}\),
          \(\displaystyle {100}\)
        },
        xtick={0,0.2,0.4,0.6,0.8,1},
        xticklabels={
          \(\displaystyle {0}\),
          \(\displaystyle {20}\),
          \(\displaystyle {40}\),
          \(\displaystyle {60}\),
          \(\displaystyle {80}\),
          \(\displaystyle {100}\)
        }
      ]
      \addplot [line width=1.2, pscc_purple]
      table {%
        0.357142925262451 0
        0.631578922271729 0
        0.709677457809448 0.00155520439147949
        0.709677457809448 0.00311040878295898
        0.730769157409668 0.00622081756591797
        0.74193549156189 0.00777602195739746
        0.74193549156189 0.017107367515564
        0.769230842590332 0.0186625719070435
        0.769230842590332 0.0248833894729614
        0.774193525314331 0.0264385938644409
        0.774193525314331 0.0373250246047974
        0.785714268684387 0.0388802289962769
        0.799999952316284 0.0404354333877563
        0.799999952316284 0.0482114553451538
        0.806451559066772 0.0497667789459229
        0.806451559066772 0.0855365991592407
        0.807692289352417 0.0870918035507202
        0.807692289352417 0.0964230298995972
        0.818181753158569 0.0979782342910767
        0.82608699798584 0.0995334386825562
        0.82608699798584 0.101088643074036
        0.82758617401123 0.102643847465515
        0.838709592819214 0.105754256248474
        0.838709592819214 0.183514833450317
        0.839999914169312 0.185070037841797
        0.839999914169312 0.199066877365112
        0.842105269432068 0.200622081756592
        0.846153855323792 0.202177286148071
        0.846153855323792 0.23483669757843
        0.857142925262451 0.23639190196991
        0.857142925262451 0.239502310752869
        0.866666674613953 0.241057515144348
        0.86956524848938 0.242612719535828
        0.870967745780945 0.244167923927307
        0.870967745780945 0.35303258895874
        0.879999995231628 0.354587912559509
        0.879999995231628 0.373250365257263
        0.884615421295166 0.374805569648743
        0.884615421295166 0.407464981079102
        0.888888835906982 0.409020185470581
        0.896551728248596 0.410575389862061
        0.896551728248596 0.41213059425354
        0.899999976158142 0.41368579864502
        0.899999976158142 0.415241003036499
        0.903225779533386 0.416796207427979
        0.903225779533386 0.578538179397583
        0.904761910438538 0.580093383789062
        0.91304349899292 0.581648588180542
        0.920000076293945 0.583203792572021
        0.920000076293945 0.603421449661255
        0.923076868057251 0.604976654052734
        0.923076868057251 0.643857002258301
        0.928571462631226 0.64541220664978
        0.931034564971924 0.64696741104126
        0.935483932495117 0.648522615432739
        0.935483932495117 0.780715465545654
        0.941176414489746 0.782270669937134
        0.950000047683716 0.783825874328613
        0.950000047683716 0.788491487503052
        0.960000038146973 0.790046691894531
        0.960000038146973 0.800933122634888
        0.961538434028625 0.802488327026367
        0.961538434028625 0.832037329673767
        0.967741966247559 0.833592534065247
        0.967741966247559 0.925349950790405
        1 0.926905155181885
        1 1
      };
      \addlegendentry{ST, CP}
      \addplot [line width=1.2, pscc_orange]
      table {%
        0.357142925262451 0
        0.631578922271729 0.00155520439147949
        0.677419424057007 0.00311040878295898
        0.677419424057007 0.00466561317443848
        0.709677457809448 0.00622081756591797
        0.709677457809448 0.00777602195739746
        0.720000028610229 0.00933122634887695
        0.74193549156189 0.0108864307403564
        0.74193549156189 0.0155520439147949
        0.759999990463257 0.017107367515564
        0.759999990463257 0.0186625719070435
        0.761904716491699 0.0202177762985229
        0.769230842590332 0.0217729806900024
        0.769230842590332 0.0233281850814819
        0.774193525314331 0.0248833894729614
        0.774193525314331 0.0404354333877563
        0.78947377204895 0.0419906377792358
        0.799999952316284 0.0435458421707153
        0.799999952316284 0.0482114553451538
        0.806451559066772 0.0497667789459229
        0.806451559066772 0.0886470079421997
        0.807692289352417 0.0902022123336792
        0.807692289352417 0.104199051856995
        0.809523820877075 0.105754256248474
        0.818181753158569 0.107309460639954
        0.82608699798584 0.108864665031433
        0.82758617401123 0.110419869422913
        0.838709592819214 0.111975073814392
        0.838709592819214 0.186625242233276
        0.839999914169312 0.188180446624756
        0.839999914169312 0.203732490539551
        0.846153855323792 0.20528769493103
        0.846153855323792 0.241057515144348
        0.857142925262451 0.242612719535828
        0.857142925262451 0.244167923927307
        0.866666674613953 0.245723128318787
        0.866666674613953 0.247278332710266
        0.86956524848938 0.248833656311035
        0.86956524848938 0.250388860702515
        0.870967745780945 0.251944065093994
        0.870967745780945 0.357698321342468
        0.879999995231628 0.359253525733948
        0.879999995231628 0.373250365257263
        0.884615421295166 0.374805569648743
        0.884615421295166 0.399688959121704
        0.896551728248596 0.401244163513184
        0.899999976158142 0.402799367904663
        0.899999976158142 0.405909776687622
        0.903225779533386 0.407464981079102
        0.903225779533386 0.566096425056458
        0.909090995788574 0.567651629447937
        0.91304349899292 0.569206833839417
        0.920000076293945 0.570762038230896
        0.920000076293945 0.597200632095337
        0.923076868057251 0.598755836486816
        0.923076868057251 0.640746474266052
        0.928571462631226 0.642301797866821
        0.931034564971924 0.643857002258301
        0.931034564971924 0.64541220664978
        0.935483932495117 0.64696741104126
        0.935483932495117 0.797822713851929
        0.944444417953491 0.799377918243408
        0.950000047683716 0.800933122634888
        0.950000047683716 0.802488327026367
        0.960000038146973 0.804043531417847
        0.960000038146973 0.814929962158203
        0.961538434028625 0.816485166549683
        0.961538434028625 0.846034288406372
        0.967741966247559 0.847589492797852
        0.967741966247559 0.930015563964844
        1 0.931570768356323
        1 1
      };
      \addlegendentry{ST, NP}
      \addplot [line width=1.2, pscc_green]
      table {%
        0.357142925262451 0
        0.631578922271729 0
        0.709677457809448 0.00155520439147949
        0.709677457809448 0.00466561317443848
        0.727272748947144 0.00622081756591797
        0.74193549156189 0.00777602195739746
        0.74193549156189 0.0124416351318359
        0.75 0.0139968395233154
        0.759999990463257 0.0155520439147949
        0.759999990463257 0.017107367515564
        0.769230842590332 0.0186625719070435
        0.769230842590332 0.0248833894729614
        0.774193525314331 0.0264385938644409
        0.774193525314331 0.0311042070388794
        0.78947377204895 0.0326594114303589
        0.793103456497192 0.0342146158218384
        0.799999952316284 0.0357698202133179
        0.799999952316284 0.0388802289962769
        0.806451559066772 0.0404354333877563
        0.806451559066772 0.0917574167251587
        0.807692289352417 0.0933126211166382
        0.807692289352417 0.107309460639954
        0.82608699798584 0.108864665031433
        0.833333253860474 0.110419869422913
        0.833333253860474 0.111975073814392
        0.838709592819214 0.113530278205872
        0.838709592819214 0.186625242233276
        0.839999914169312 0.188180446624756
        0.839999914169312 0.192846059799194
        0.846153855323792 0.194401264190674
        0.846153855323792 0.222395062446594
        0.850000023841858 0.223950266838074
        0.857142925262451 0.225505471229553
        0.857142925262451 0.228615880012512
        0.866666674613953 0.230171084403992
        0.86956524848938 0.231726288795471
        0.86956524848938 0.233281493186951
        0.870967745780945 0.23483669757843
        0.870967745780945 0.362363934516907
        0.879999995231628 0.363919138908386
        0.879999995231628 0.393468141555786
        0.884615421295166 0.395023345947266
        0.884615421295166 0.435458779335022
        0.892857074737549 0.438569188117981
        0.899999976158142 0.44167959690094
        0.903225779533386 0.443234801292419
        0.903225779533386 0.590979814529419
        0.904761910438538 0.592535018920898
        0.904761910438538 0.594090223312378
        0.909090995788574 0.595645427703857
        0.91304349899292 0.597200632095337
        0.920000076293945 0.598755836486816
        0.920000076293945 0.620528697967529
        0.923076868057251 0.622083902359009
        0.923076868057251 0.654743432998657
        0.931034564971924 0.656298637390137
        0.931034564971924 0.657853841781616
        0.935483932495117 0.659409046173096
        0.935483932495117 0.76516330242157
        0.941176414489746 0.766718506813049
        0.950000047683716 0.768273711204529
        0.950000047683716 0.772939324378967
        0.960000038146973 0.774494528770447
        0.960000038146973 0.786936283111572
        0.961538434028625 0.788491487503052
        0.961538434028625 0.821150779724121
        0.964285731315613 0.822705984115601
        0.967741966247559 0.82426118850708
        0.967741966247559 0.944012403488159
        1 0.945567607879639
        1 1
      };
      \addlegendentry{MT, CP}
      \addplot [line width=1.2, pscc_blue]
      table {%
        0.357142925262451 0
        0.533333301544189 0.00155520439147949
        0.538461565971375 0.00311040878295898
        0.576923131942749 0.00466561317443848
        0.576923131942749 0.00933122634887695
        0.580645084381104 0.0108864307403564
        0.580645084381104 0.0124416351318359
        0.600000023841858 0.0139968395233154
        0.600000023841858 0.0202177762985229
        0.612903237342834 0.0217729806900024
        0.612903237342834 0.0279937982559204
        0.615384578704834 0.0295490026473999
        0.639999985694885 0.0311042070388794
        0.639999985694885 0.0326594114303589
        0.642857074737549 0.0342146158218384
        0.645161271095276 0.0357698202133179
        0.645161271095276 0.0419906377792358
        0.653846144676208 0.0435458421707153
        0.653846144676208 0.0451010465621948
        0.666666746139526 0.0466562509536743
        0.666666746139526 0.0482114553451538
        0.677419424057007 0.0497667789459229
        0.677419424057007 0.0653188228607178
        0.680000066757202 0.0668740272521973
        0.692307710647583 0.0684292316436768
        0.692307710647583 0.0793156623840332
        0.709677457809448 0.0808708667755127
        0.709677457809448 0.1181960105896
        0.720000028610229 0.119751214981079
        0.720000028610229 0.122861623764038
        0.730769157409668 0.124416828155518
        0.730769157409668 0.139968872070312
        0.74193549156189 0.141524076461792
        0.74193549156189 0.216174125671387
        0.75 0.217729330062866
        0.75 0.219284653663635
        0.758620738983154 0.220839858055115
        0.759999990463257 0.222395062446594
        0.759999990463257 0.228615880012512
        0.769230842590332 0.230171084403992
        0.769230842590332 0.247278332710266
        0.772727251052856 0.248833656311035
        0.774193525314331 0.250388860702515
        0.774193525314331 0.335925340652466
        0.785714268684387 0.337480545043945
        0.793103456497192 0.340590953826904
        0.799999952316284 0.342146158218384
        0.799999952316284 0.360808730125427
        0.806451559066772 0.362363934516907
        0.806451559066772 0.452566146850586
        0.807692289352417 0.454121351242065
        0.807692289352417 0.480559825897217
        0.809523820877075 0.482115030288696
        0.821428537368774 0.483670234680176
        0.82608699798584 0.485225439071655
        0.82608699798584 0.486780643463135
        0.838709592819214 0.488335967063904
        0.838709592819214 0.573872447013855
        0.839999914169312 0.575427770614624
        0.839999914169312 0.584758996963501
        0.842105269432068 0.58631420135498
        0.846153855323792 0.58786940574646
        0.846153855323792 0.612752676010132
        0.850000023841858 0.614307880401611
        0.862068891525269 0.615863084793091
        0.863636374473572 0.61741828918457
        0.86956524848938 0.61897349357605
        0.870967745780945 0.620528697967529
        0.870967745780945 0.704510092735291
        0.879999995231628 0.70606529712677
        0.879999995231628 0.721617460250854
        0.884615421295166 0.723172664642334
        0.884615421295166 0.755831956863403
        0.892857074737549 0.758942484855652
        0.899999976158142 0.762052893638611
        0.899999976158142 0.76360809803009
        0.903225779533386 0.76516330242157
        0.903225779533386 0.860031127929688
        0.920000076293945 0.861586332321167
        0.920000076293945 0.869362354278564
        0.923076868057251 0.870917558670044
        0.923076868057251 0.886469602584839
        0.928571462631226 0.888024806976318
        0.935483932495117 0.889580011367798
        0.935483932495117 0.953343629837036
        0.950000047683716 0.954898834228516
        0.950000047683716 0.956454038619995
        0.95652174949646 0.958009243011475
        0.960000038146973 0.959564566612244
        0.960000038146973 0.961119771003723
        0.961538434028625 0.962674975395203
        0.961538434028625 0.968895792961121
        0.967741966247559 0.9704509973526
        0.967741966247559 0.987558364868164
        1 0.989113569259644
        1 1
      };
      \addlegendentry{MT, NP}
      \addplot [line width=1.2, black, dashed]
      table {%
        0.899999976158142 -0.05
        0.899999976158142 1.05
      };
      % \addlegendentry{Perfect Percentiles}
      \coordinate (spy point) at (axis cs:0.625,0.26);
      \coordinate (spy view) at (axis cs:0.2,0.16);
    \end{axis}
  \end{tikzpicture}
};
\spy on (spy point) in node [fill=white] at (spy view);
\end{tikzpicture}

%% file: fig_Highest_Coverage_Transformer.tex
% This file was created with matplot2tikz v0.3.2.
\begin{tikzpicture}

\definecolor{darkgray176}{RGB}{176,176,176}
\definecolor{gray}{RGB}{128,128,128}
\definecolor{lightgray204}{RGB}{204,204,204}
\definecolor{steelblue31119180}{RGB}{31,119,180}

\begin{axis}[
legend cell align={left},
legend columns = 3,
label style={font=\footnotesize},
legend style={fill opacity=0.8, draw opacity=1, text opacity=1, draw=lightgray204, font=\footnotesize, at={(0.5,-0.22)}, anchor=north, draw=none},
tick align=inside,
tick pos=both,
x grid style={darkgray176},
xlabel={Date},
xmajorgrids,
width = \columnwidth,
height = 0.65\columnwidth,
tick label style={font=\footnotesize},
x tick label style={rotate=0},
xmin=20118.5, xmax=20151.5,
xtick style={color=black, font=\footnotesize},
xtick={20120,20124,20128,20132,20136,20140,20144,20148},
xticklabels={
  02-01,
  02-05,
  02-09,
  02-13,
  02-17,
  02-21,
  02-25,
  03-01
},
scaled x ticks = false,
y grid style={darkgray176},
ylabel={Scale Factor (k)},
ymajorgrids,
ymin=1.12058877609764, ymax=1.29433847317561,
ytick style={color=black},
ytick={1.1,1.15,1.2,1.25},
yticklabels={
  \(\displaystyle {1.10}\),
  \(\displaystyle {1.15}\),
  \(\displaystyle {1.20}\),
  \(\displaystyle {1.25}\)
},
]
\path [draw=gray, fill=gray, opacity=0.5]
(axis cs:20120,1.25946840582674)
--(axis cs:20120,1.18656623600293)
--(axis cs:20121,1.18654402206196)
--(axis cs:20122,1.1746754722977)
--(axis cs:20123,1.16883746350444)
--(axis cs:20124,1.19004697451557)
--(axis cs:20125,1.19410554470633)
--(axis cs:20126,1.19756604061995)
--(axis cs:20127,1.19450704652738)
--(axis cs:20128,1.18706474843207)
--(axis cs:20129,1.19535709166382)
--(axis cs:20130,1.19672430865212)
--(axis cs:20131,1.19285195159933)
--(axis cs:20132,1.196866936087)
--(axis cs:20133,1.1953767147669)
--(axis cs:20134,1.19690454836832)
--(axis cs:20135,1.19833571021489)
--(axis cs:20136,1.19786708187209)
--(axis cs:20137,1.19313105672185)
--(axis cs:20138,1.17299741725287)
--(axis cs:20139,1.12848648960119)
--(axis cs:20140,1.13818651299269)
--(axis cs:20141,1.15101199259337)
--(axis cs:20142,1.14951885477632)
--(axis cs:20143,1.15278379853412)
--(axis cs:20144,1.17602648599167)
--(axis cs:20145,1.1816824088315)
--(axis cs:20146,1.18269058449999)
--(axis cs:20147,1.18150681969055)
--(axis cs:20148,1.18083754461168)
--(axis cs:20149,1.18842862992245)
--(axis cs:20150,1.16966547954437)
--(axis cs:20150,1.24064677429075)
--(axis cs:20150,1.24064677429075)
--(axis cs:20149,1.26018962190127)
--(axis cs:20148,1.24772480811217)
--(axis cs:20147,1.25026552187868)
--(axis cs:20146,1.24752687046693)
--(axis cs:20145,1.25231442877631)
--(axis cs:20144,1.24217351147352)
--(axis cs:20143,1.22668413503857)
--(axis cs:20142,1.22393582082865)
--(axis cs:20141,1.22702354543129)
--(axis cs:20140,1.20996776597654)
--(axis cs:20139,1.20833132888679)
--(axis cs:20138,1.24884524726175)
--(axis cs:20137,1.26496965184821)
--(axis cs:20136,1.28268078726033)
--(axis cs:20135,1.28587282615583)
--(axis cs:20134,1.26985212754229)
--(axis cs:20133,1.27194645515415)
--(axis cs:20132,1.27643974459501)
--(axis cs:20131,1.2670884597122)
--(axis cs:20130,1.26793397208879)
--(axis cs:20129,1.27209999999872)
--(axis cs:20128,1.25626868276363)
--(axis cs:20127,1.27499255371749)
--(axis cs:20126,1.28644075967207)
--(axis cs:20125,1.27415735233448)
--(axis cs:20124,1.25844063940788)
--(axis cs:20123,1.23856745361466)
--(axis cs:20122,1.24840630702258)
--(axis cs:20121,1.26389447984769)
--(axis cs:20120,1.25946840582674)
--cycle;
\addlegendimage{area legend, draw=gray, fill=gray, opacity=0.5}
\addlegendentry{5th-95th Percentile}

\addplot [line width=1.2, black, dashed]
table {%
20120 1.23503484442642
20121 1.22155994310951
20122 1.20761137492745
20123 1.19903721968632
20124 1.2251186971554
20125 1.22610167797354
20126 1.24182254380344
20127 1.22233006692545
20128 1.2336595012786
20129 1.23280577711313
20130 1.2191180188086
20131 1.20418417399433
20132 1.22469068709184
20133 1.23251161599098
20134 1.26126477196287
20135 1.28195308663204
20136 1.22105292499975
20137 1.21859410338778
20138 1.19070927267741
20139 1.13615190314431
20140 1.1400550657692
20141 1.17959499152689
20142 1.15894002738042
20143 1.16379052545955
20144 1.20683895528286
20145 1.19885541811466
20146 1.19092212966503
20147 1.19800399016419
20148 1.21108652880203
20149 1.23126552948124
20150 1.18005405633799
};
\addlegendentry{True k}
\addplot [line width=1.2, pscc_purple]
table {%
20120 1.23496458504689
20121 1.22823678186127
20122 1.21112480250663
20123 1.20653478782143
20124 1.22997436836589
20125 1.24451763089605
20126 1.26355604026139
20127 1.24226089181883
20128 1.23374076091253
20129 1.24710468028496
20130 1.23841392925853
20131 1.23446672119092
20132 1.24004547570511
20133 1.23966748438112
20134 1.24010012025634
20135 1.25087911895043
20136 1.24832913865713
20137 1.23686493110336
20138 1.20094745329871
20139 1.14964995494189
20140 1.15674890372113
20141 1.18345691096207
20142 1.17375175918709
20143 1.17247466757487
20144 1.20550823148177
20145 1.21990789515743
20146 1.21224473634517
20147 1.21004793100343
20148 1.21123604639684
20149 1.22565850285614
20150 1.19217128653373
};
\addlegendentry{Predicted k (50th Percentile)}
\end{axis}

\end{tikzpicture}

%% file: fig_Lowest_Coverage_Transformer.tex
% This file was created with matplot2tikz v0.3.2.
\begin{tikzpicture}

\definecolor{darkgray176}{RGB}{176,176,176}
\definecolor{gray}{RGB}{128,128,128}
\definecolor{lightgray204}{RGB}{204,204,204}
\definecolor{steelblue31119180}{RGB}{31,119,180}

\begin{axis}[
legend cell align={left},
legend columns = 3,
label style={font=\footnotesize},
legend style={fill opacity=0.8, draw opacity=1, text opacity=1, draw=lightgray204, font=\footnotesize, at={(0.5,-0.22)}, anchor=north, draw=none},
tick align=inside,
tick pos=both,
x grid style={darkgray176},
xlabel={Date},
xmajorgrids,
width = \columnwidth,
height = 0.65\columnwidth,
tick label style={font=\footnotesize},
x tick label style={rotate=0},
xmin=20118.5, xmax=20151.5,
xtick style={color=black, font=\footnotesize},
xtick={20120,20124,20128,20132,20136,20140,20144,20148},
xticklabels={
  02-01,
  02-05,
  02-09,
  02-13,
  02-17,
  02-21,
  02-25,
  03-01
},
scaled x ticks = false,
y grid style={darkgray176},
ylabel={Scale Factor (k)},
ymajorgrids,
ymin=1.07591116263133, ymax=1.27492679818455,
ytick={1.1,1.15,1.2,1.25},
yticklabels={
  \(\displaystyle {1.10}\),
  \(\displaystyle {1.15}\),
  \(\displaystyle {1.20}\),
  \(\displaystyle {1.25}\)
},
]
\path [draw=gray, fill=gray, opacity=0.5]
(axis cs:20120,1.2449511844142)
--(axis cs:20120,1.16698438708787)
--(axis cs:20121,1.17374992109207)
--(axis cs:20122,1.14574159714993)
--(axis cs:20123,1.137972831306)
--(axis cs:20124,1.17394103812906)
--(axis cs:20125,1.18574719585827)
--(axis cs:20126,1.19168753898312)
--(axis cs:20127,1.18039287617439)
--(axis cs:20128,1.16196977643329)
--(axis cs:20129,1.18097568608932)
--(axis cs:20130,1.17772914896411)
--(axis cs:20131,1.17232629642248)
--(axis cs:20132,1.18688899049556)
--(axis cs:20133,1.17932262354054)
--(axis cs:20134,1.17499912501988)
--(axis cs:20135,1.1857630955467)
--(axis cs:20136,1.18447714373624)
--(axis cs:20137,1.17623210563761)
--(axis cs:20138,1.14533565664766)
--(axis cs:20139,1.09128252520555)
--(axis cs:20140,1.09166043913518)
--(axis cs:20141,1.1195331771346)
--(axis cs:20142,1.11794228547199)
--(axis cs:20143,1.12238923999997)
--(axis cs:20144,1.13911894004632)
--(axis cs:20145,1.14996062182636)
--(axis cs:20146,1.15097804875441)
--(axis cs:20147,1.15010442515555)
--(axis cs:20148,1.15330506998367)
--(axis cs:20149,1.15545706554051)
--(axis cs:20150,1.13054299637695)
--(axis cs:20150,1.22465277234787)
--(axis cs:20150,1.22465277234787)
--(axis cs:20149,1.24365861294657)
--(axis cs:20148,1.24235526076226)
--(axis cs:20147,1.23460217367935)
--(axis cs:20146,1.22928805353571)
--(axis cs:20145,1.23517737833656)
--(axis cs:20144,1.23073616159227)
--(axis cs:20143,1.21391234535588)
--(axis cs:20142,1.20950398470047)
--(axis cs:20141,1.21013045302764)
--(axis cs:20140,1.20078217924595)
--(axis cs:20139,1.20058561549475)
--(axis cs:20138,1.2338898593853)
--(axis cs:20137,1.25292488019989)
--(axis cs:20136,1.26165113458103)
--(axis cs:20135,1.25608377717487)
--(axis cs:20134,1.25892337842192)
--(axis cs:20133,1.25831307975708)
--(axis cs:20132,1.26588063293213)
--(axis cs:20131,1.2449524734841)
--(axis cs:20130,1.26523552862256)
--(axis cs:20129,1.25559330113722)
--(axis cs:20128,1.24189981388141)
--(axis cs:20127,1.25131885938636)
--(axis cs:20126,1.26381107806749)
--(axis cs:20125,1.25420422415424)
--(axis cs:20124,1.25589473632135)
--(axis cs:20123,1.22311739885602)
--(axis cs:20122,1.23103916377631)
--(axis cs:20121,1.24131950000109)
--(axis cs:20120,1.2449511844142)
--cycle;
\addlegendimage{area legend, draw=gray, fill=gray, opacity=0.5}
\addlegendentry{5th-95th Percentile}

\addplot [line width=1.2, black, dashed]
table {%
20120 1.18456243975002
20121 1.21623348318638
20122 1.21297118630257
20123 1.17585673199767
20124 1.17387387241619
20125 1.24924881414332
20126 1.20177330510937
20127 1.19620063682613
20128 1.17220207190642
20129 1.25877259894339
20130 1.17654555167908
20131 1.21005329926458
20132 1.22095472358918
20133 1.2455105546141
20134 1.16726797451284
20135 1.21620013181388
20136 1.18434958275924
20137 1.2074798220724
20138 1.19739647461731
20139 1.12456395389684
20140 1.08495732788375
20141 1.14700537351837
20142 1.14384127919036
20143 1.13443214868399
20144 1.13310475802004
20145 1.15766059034798
20146 1.22044540945286
20147 1.20241417182379
20148 1.24379080177584
20149 1.15692381702483
20150 1.16100419131793
};
\addlegendentry{True k}
\addplot [line width=1.2, pscc_purple]
table {%
20120 1.1945226017191
20121 1.20264620412089
20122 1.17324323915461
20123 1.17109130834339
20124 1.20639265613098
20125 1.21936485609002
20126 1.2185927703718
20127 1.20726812652766
20128 1.19613144931324
20129 1.20743855049592
20130 1.20841468765056
20131 1.20243445873872
20132 1.22085930987127
20133 1.21400488926374
20134 1.2071464305686
20135 1.21605253519895
20136 1.21982402670099
20137 1.20882905461736
20138 1.17965957909826
20139 1.12359803242848
20140 1.12641809837513
20141 1.14866853781803
20142 1.14556967228395
20143 1.15205442285559
20144 1.17235173311729
20145 1.18821573118243
20146 1.19145780047364
20147 1.19324835947232
20148 1.19723456526816
20149 1.1974652004018
20150 1.15991403068368
};
\addlegendentry{Predicted k (50th Percentile)}
\end{axis}

\end{tikzpicture}

%% file: fig_load_direct_histogram.tex
\definecolor{darkgray176}{RGB}{176,176,176}
\definecolor{darkorange25512714}{RGB}{255,127,14}
\definecolor{steelblue31119180}{RGB}{31,119,180}

\begin{tikzpicture}
\begin{axis}[
    ybar, % Apply ybar globally
    legend cell align={left},
    legend columns = 2,
    label style={font=\footnotesize},
    legend style={fill opacity=0.8, draw opacity=1, text opacity=1, font=\footnotesize, at={(0.5,0.95)}, anchor=north, draw=none},
    bar width=0.05, % matches bin width
    ymin=0,
    ymax=250,
    xlabel={Coverage (\%)},
    ylabel={Transformer Count},
    xtick={0,0.1,0.2,0.3,0.4,0.5,0.6,0.7,0.8,0.9,1},
    xticklabels={
        \(\displaystyle{0}\),
        \(\displaystyle{10}\),
        \(\displaystyle{20}\),
        \(\displaystyle{30}\),
        \(\displaystyle{40}\),
        \(\displaystyle{50}\),
        \(\displaystyle{60}\),
        \(\displaystyle{70}\),
        \(\displaystyle{80}\),
        \(\displaystyle{90}\),
        \(\displaystyle{100}\),
        }, % bin edges
    xmin=0,
    xmax=1,
    ytick = {0,50,100,150,200,250},
    tick align=inside,
    tick pos=both,
    x grid style={darkgray176},
    xmajorgrids,
    tick label style={font=\footnotesize},
    xtick style={color=black, font=\footnotesize},
    y grid style={darkgray176},
    ymajorgrids,
    ytick style={color=black},
    label style={font=\footnotesize},
    width = \columnwidth,
    height = 0.65\columnwidth,
    legend image code/.code={%
        \draw[#1,fill=#1] (0cm,-0.1cm) rectangle (0.3cm,0.1cm); % Custom legend marker
    },
]
% Shift bars by half a bin so they sit between ticks
\addplot+[
    bar shift=0.025,
    fill=pscc_orange, % Set bar color
    opacity=0.7,
    color = pscc_orange, % Set border color
] coordinates {
    (0.00, 196.0)
    (0.05, 96.0)
    (0.10, 61.0)
    (0.15, 74.0)
    (0.20, 44.0)
    (0.25, 48.0)
    (0.30, 22.0)
    (0.35, 32.0)
    (0.40, 16.0)
    (0.45, 18.0)
    (0.50, 13.0)
    (0.55, 4.0)
    (0.60, 8.0)
    (0.65, 2.0)
    (0.70, 6.0)
    (0.75, 1.0)
    (0.80, 1.0)
    (0.85, 1.0)
    (0.90, 0.0)
    (0.95, 1.0)
};
\addlegendentry{Load Approach}
\addplot+[
    bar shift=0.025,
    fill=pscc_blue, % Set bar color
    opacity=0.7,
    color=pscc_blue,
] coordinates {
    (0.00, 0.0)
    (0.05, 0.0)
    (0.10, 0.0)
    (0.15, 0.0)
    (0.20, 0.0)
    (0.25, 0.0)
    (0.30, 0.0)
    (0.35, 0.0)
    (0.40, 0.0)
    (0.45, 0.0)
    (0.50, 0.0)
    (0.55, 0.0)
    (0.60, 1.0)
    (0.65, 0.0)
    (0.70, 11.0)
    (0.75, 14.0)
    (0.80, 126.0)
    (0.85, 114.0)
    (0.90, 242.0)
    (0.95, 136.0)
};
\addlegendentry{Direct Approach}
\end{axis}
\end{tikzpicture}

%% file: conclusions.tex
This paper introduced a probabilistic forecasting framework to assign relay scale factors, enable the automated application of DTR for distribution transformers.
The approach uses clustering to enable improved forecast accuracy across the fleet of transformers, using retrospective optimal scale factors as training labels for the forecasting model.
This direct prediction of scale factors achieves reliable risk quantification, whilst it is shown that an alternative load prediction-based approach severely underestimates the risk.
This result was shown to be robust even in the presence of realistic temperature forecast errors.
Future works could study multi-year datasets and explore more frequent relay setting updates, which could further increase the capacity that can be released. The techno-economic implications of implementing and maintaining this approach across a fleet of transformers and the cumulative ageing effects of more frequent operation of transformers near their thermal limits could also be explored.

In summary, direct prediction of the scale factor of Adaptive PDs is shown to provide an effective, automated approach that could be used to achieve risk-aware DTR. It is concluded that the proposed approach shows strong promise to enable DNOs to provide a fleet-wide automation approach that navigates the trade-off between capacity gain from DTR and asset risk.